%% file: Master.tex
\documentclass[11pt]{article}

\usepackage[title,toc,titletoc]{appendix}
\usepackage{etoolbox}
\usepackage{graphicx} 
\usepackage{subfiles}
\usepackage{float}
\usepackage[a4paper, total={6in, 8in}]{geometry}
\usepackage{color,soul}
\usepackage{amsthm}
\usepackage{amsmath,amssymb,tabu}
\usepackage{amsfonts}
\usepackage{enumitem}
\usepackage[english]{babel}
\usepackage{mathrsfs}  
\usepackage{mathtools}
\usepackage[mathscr]{euscript}
\usepackage{subfig}
\usepackage{natbib}
\usepackage{adjustbox}
\usepackage{multirow}
\usepackage{xr}
\usepackage{thmtools}
\usepackage{hyperref}
\usepackage[capitalize]{cleveref}

\patchcmd{\appendices}{\quad}{. }{}{}
\def\bs{\ensuremath\boldsymbol}

\def\bf{\textbf}

\DeclareMathOperator*{\argmin}{arg\,min}

\newtheorem{proposition}{\textbf{{Proposition}}}
\newtheorem{corollary}{Corollary}
\newtheorem{theorem}{Theorem}
\theoremstyle{definition}
\newtheorem{remark}{Remark}

\newtheorem*{AssumptionMD}{\textbf{{Assumption MD}}} 
\newtheorem*{AssumptionER}{\textbf{{Assumption ER}}}  
\newtheorem*{AssumptionCS}{\textbf{{Assumption CS}}}
\newtheorem*{AssumptionAE}{\textbf{{Assumption AE}}} 
\newtheorem*{AssumptionER*}{\textbf{{Assumption ER*}}}  
\newtheorem*{AssumptionAD}{\textbf{{Assumption AD}}}
\newtheorem*{AssumptionBE}{\textbf{{Assumption BE}}} 
\newtheorem*{AssumptionSN}{\textbf{{Assumption SN}}} 
\newtheorem*{AssumptionLR}{\textbf{{Assumption LR}}}  

\def\tx{\text}
\newcommand\smallO{
	\mathchoice
	{{\scriptstyle\mathcal{O}}}
	{{\scriptstyle\mathcal{O}}}
	{{\scriptscriptstyle\mathcal{O}}}
	{\scalebox{.7}{$\scriptscriptstyle\mathcal{O}$}}%
}
\newenvironment{pfff}[1][Proof:]{\vspace{1ex}{\noindent\textbf{#1}}}
{\hfill\qed\vspace{1ex}}

\makeatletter
\def\blfootnote{\xdef\@thefnmark{}\@footnotetext}
\makeatother

\numberwithin{equation}{section}

\declaretheoremstyle[bodyfont=\slshape]{slshape}

\newlist{thmlist}{enumerate}{1}
\setlist[thmlist]{label=\textup{(\roman{thmlisti})},
	ref=\thethm\textup{(\roman{thmlisti})}}

\Crefname{thm}{Theorem}{Theorems}
\Crefname{lem}{Lemma}{Lemmas}
\Crefname{asm}{Assumption}{Assumptions}
\Crefname{listthm}{Theorem}{Theorems}
\Crefname{listlem}{Lemma}{Lemmas}
\Crefname{listasm}{Assumption}{Assumptions}

\addtotheorempostheadhook[thm]{\crefalias{thmlisti}{listthm}}
\addtotheorempostheadhook[lem]{\crefalias{thmlisti}{listlem}}
\addtotheorempostheadhook[asm]{\crefalias{asmlisti}{listasm}}

\begin{document}

\title{Fixed $T$ Estimation of Linear Panel Data Models with Interactive Fixed Effects$^*$}
\author{Ayden Higgins$^{\dagger}$\\University of Cambridge}
\date{\today}
\maketitle

\begin{abstract}
This paper studies the estimation of linear panel data models with interactive fixed effects, where one dimension of the panel, typically time, may be fixed. To this end, a novel transformation is introduced that reduces the model to a lower dimension, and, in doing so, relieves the model of incidental parameters in the cross-section. The central result of this paper demonstrates that transforming the model and then applying the principal component (PC) estimator of \cite{bai_panel_2009} delivers $\sqrt{n}$ consistent estimates of regression slope coefficients with $T$ fixed. Moreover, these estimates are shown to be asymptotically unbiased in the presence of cross-sectional dependence, serial dependence, and with the inclusion of dynamic regressors, in stark contrast to the usual case. The large $n$, large $T$ properties of this approach are also studied, where many of these results carry over to the case in which $n$ is growing sufficiently fast relative to $T$. Transforming the model also proves to be useful beyond estimation, a point illustrated by showing that with $T$ fixed, the eigenvalue ratio test of \cite{horenstein} provides a consistent test for the number of factors when applied to the transformed model. \\

\color{white}.\color{black}
\end{abstract}

\vspace{-0.1in}
\blfootnote{\hspace{-0.2in}$^*$This work was supported by the European Research Council through the grant ERC-2016-STG-715787-MiMo. I am grateful to Valentina Corradi, Koen Jochmans, Federico Martellosio, Jo$\tilde{\text{a}}$o Santos Silva and Martin Weidner for their comments and suggestions.}
\blfootnote{\hspace{-0.2in}$^{\dagger}$Email: amh239@cam.ac.uk.}
\newpage


\subfile{Body/MainText.tex}

\newpage


\end{document}

%% file: Body/MainText.tex
\section{Introduction}
This paper contributes to the extensive literature on linear panel data models with interactive effects. These models have proven to be very popular since, in many situations, the existence of such structures is well motivated; for example, arising due to unobserved heterogeneity across individuals, or exposure to common shocks. The model studied in this paper assumes that, in a panel with entries indexed $i = 1,...,n$, $t = 1,...,T$, outcomes are generated according to  
\begin{align}
	\bs{y}_{t} 
		= \alpha \bs{y}_{t-1} + \bs{X}_t \bs{\beta} + \bs{\Lambda}^* \bs{f}^*_t + \bs{\varepsilon}_t, \label{eq1}
\end{align}
where $\bs{y}_t$ and $\bs{\varepsilon}_t$ are $n \times 1$ vectors of outcomes and error terms, respectively, $\bs{X}_t$ is an $n \times K$ matrix of exogenous covariates, $\bs{\Lambda}^*$ is an $n \times R^*$ matrix of time-invariant factor loadings, and $\bs{f}_t^*$ is an $R^* \times 1$ vector of time-varying factors. It is assumed that both the outcomes and the covariates are observed by the econometrician, while the factors, the loadings, and the error terms are not. The parameter of interest in this model is the $(K+1) \times 1$ vector $\bs{\theta} \coloneqq (\alpha,\bs{\beta}^\top)^\top$, comprised of the scalar autoregressive parameter $\alpha$ and the $K \times 1$ vector $\bs{\beta}$.

This model can be seen as a generalisation of familiar models of additive effects, such as individual, time or group effects. For example, individual and time effects nest as a special case of \eqref{eq1} in which 
\begin{align}
	\bs{\Lambda}^*
	= 
	\begin{pmatrix}
		\lambda_1      & 1      \\
		\vdots         & \vdots \\
		\lambda_{n}    & 1 
	\end{pmatrix},\ 
	\bs{f}_t^* 
	= 
	\begin{pmatrix}
		1 \\
		f_t 
	\end{pmatrix},
\end{align}
that is, where a vector of heterogeneous loadings is interacted with a unit factor, and where a vector of unit loadings is interacted with a time-varying factor. More generally, however, with interactive effects, no restrictions are placed on the factors or the loadings to be multiples of unit vectors, or otherwise, and both are permitted to be fully heterogeneous. 

The main obstacle to consistent estimation of $\bs{\theta}$ arises in situations where the unobserved interactive effects are somehow correlated with covariates in the model. In this event, an endogeneity problem arises resulting in standard estimators, such as least squares, producing biased estimates. One response to this is to treat the components of the factor term as additional parameters to estimate, known as the fixed effects approach. Doing this has the benefit of allowing for arbitrary correlation between the covariates, the factors and the loadings, in contrast to its main rival, random effects. However, treating both the factors and loadings as fixed effects gives rise to incidental parameters in both dimensions of the panel, which, in turn, may generate significant complications for the estimation of the parameter of interest $\bs{\theta}$, on account of the incidental parameter problem; see \cite{neyman}. Where both $n$ and $T$ are large this problem can, to some extent, be overcome, and estimation procedures have been developed for large panels which, under a broad array of circumstances, will produce consistent estimates of $\bs{\theta}$. Yet where the time dimension of the panel is small, the methods that are presently available tend to be restrictive and/or difficult to implement. 


The two main approaches currently used for short panels are the common correlated effects estimator introduced in \cite{pesaran_estimation_2006}, and the quasi-difference approach of \cite{ahn_panel_2013}. The first of these  assumes that the latent factors in the error term also impact some model covariates, such that the factors can be instrumented by cross-sectional averages. These instruments can then be levered to purge the factor term and, as a result, give rise to an estimator that is consistent with either $T$ fixed, or where both $n$ and $T$ diverge. The drawback to this approach, however, is that it relies crucially on imposing a particular functional form for the relationship between the latent factors and model covariates, which, in many instances, may not be easy to justify. The second method, the quasi-difference approach, takes advantage of the inherent indeterminacy associated with factor models. By normalising the factors and loadings in a certain way, the model can be multiplied by a difference matrix to purge the factor term. The authors apply GMM to estimate both the difference matrix and the slope coefficients, yet since the moment conditions the model yields are highly non-linear, this generates a difficult optimisation problem which, as pointed out by \cite{hayakawa}, may not satisfy the identification conditions (which are a necessary precursor to consistency) of their GMM procedure. Indeed the problem is not unique to their approach, and several closely related methods which rely on the same normalisation also suffer this affliction. 

Notwithstanding the contributions of these authors, there remains scope for a general and simple to implement method for the estimation of linear panel data models with interactive fixed effects, where the time dimension is small relative to the size of the cross-section, or indeed, may be fixed. The present paper address this by introducing a new estimation approach at the centre of which lies a transformation that relieves the model of incidental parameters in the cross-section. In contrast to other approaches, the objective of this transformation is not to purge the incidental parameters from the model entirely, but rather to transfer those in the cross-section into the time dimension, and, in doing so, facilitate the estimation of the model in situations where $T$ is small. The most appealing aspect of this transformation is its simplicity, since it is constructed directly from the data and applied to the model prior to estimation without introducing any additional parameters. Moreover, it is shown to have remarkably far-reaching consequences, and, in the main result of this paper, it is established that simply transforming the model and then applying a third estimator, the PC estimator of \cite{bai_panel_2009}, will produce $\sqrt{n}$ consistent and asymptomatically unbiased estimates with $T$ fixed, irrespective of the possible inclusion of dynamic regressors and/or the presence of cross-sectional and serial dependence in the error term. This contrasts sharply with the usual case where, with fixed $T$, the PC estimator is, in general, both inconsistent and biased outside of exceptional circumstances.

\textbf{Outline}: Section \ref{motivation} sets out the estimation approach, first introducing a transformation, and then going on to describe why the PC estimator is well suited to the task of estimating the transformed model. Section \ref{consub} begins the study of the asymptotic properties of the estimator by establishing consistency, under quite general conditions, and drawing a comparison between 
the result obtained here and those obtained when both $n$ and $T$ are large. Following this, an asymptotic expansion of the objective function is derived in Section \ref{asyexs}, from which the asymptotic distribution of the estimator is then established in Section \ref{asydist}. In order to paint a more complete picture of the estimator, multiple results are provided to cover both the situation in which $T$ is fixed, and where both $n$ and $T$ diverge. Some additional considerations are collected in Section \ref{furmat}, including estimation of the number of factors and alternative approaches to treating the initial condition. Monte Carlo simulations follow in Section \ref{sims}. Section \ref{concl} concludes. Additional discussion and proofs of the results can be found in the appendices.\color{black}

\textbf{Notation}: Throughout the paper, all vectors and matrices are real unless stated otherwise. For an $n \times 1$ vector $\bs{a}$ with elements $a_i$, $||\bs{a}||_1 \coloneqq \sum^{n}_{i=1} |a_{i}|$, $||\bs{a}||_2 \coloneqq \sqrt{\sum^{n}_{i=1} a_{i}^2}$, $|| \bs{a}||_{\infty} \coloneqq \max_{1\leq i \leq n}|a_i|$. Let $\bs{A}$ be an $n\times m$ matrix with elements $A_{ij}$. When $m =n$, and the eigenvalues of $\bs{A}$ are real, they are denoted $\mu_{\min}(\bs{A}) \coloneqq \mu_n(\bs{A}) \leq ... \leq \mu_1(\bs{A}) \eqqcolon \mu_{\max}(\bs{A})$. The following matrix norms are those induced by their vector counterparts: $||\bs{A}||_1 \coloneqq \max_{1 \leq j \leq m} \sum^{n}_{i=1} |A_{ij}|$ which is the maximum absolute column sum of $\bs{A}$, $||\bs{A}||_2 \coloneqq \sqrt{\mu_{\max}(\bs{A}^{\top}\bs{A})}$, and $||\bs{A}||_{\infty} \coloneqq \max_{1\leq i \leq n} \sum^{m}_{j=1} |A_{ij}|$ which is the maximum absolute row sum of $\bs{A}$. The Frobenius norm of $\bs{A}$ is denoted $||\bs{A}||_F \coloneqq \sqrt{\sum^{n}_{i=1} \sum^m_{j=1} A_{ij}^2} = \sqrt{\text{tr}(\bs{A}^{\top}\bs{A})}$. Let $\bs{P}_{\bs{A}} \coloneqq \bs{A}(\bs{A}^{\top }\bs{A})^{+}\bs{A}^{\top }$ and $\bs{M}_{\bs{A}} \coloneqq \bs{I}_{n} - \bs{P}_{\bs{A}}$, where $\bs{I}_n$ is the $n \times n$ identity matrix and $+$ denotes the Moore-Penrose generalised inverse. An $n \times 1$ vector of ones is denoted $\bs{\iota}_n$, and an $n \times m$ matrix of zeros is denoted $\bs{0}_{n \times m}$. The operation $\text{vec}(\cdot)$ applied to an $n\times m$ matrix $\bs{A}$ creates an $nm \times 1$ vector $\text{vec}(\bs{A})$ by stacking the columns of $\bs{A}$. The operation $\text{diag}(\cdot)$ applied to an $n \times n$ matrix $\bs{B}$ creates an $n \times n$ diagonal matrix $\text{diag}(\bs{B})$ which contains the diagonal elements of $\bs{B}$ along its diagonal, and $\text{diagv}(\bs{B})$ is used as shorthand for $\text{diag}(\bs{B})\bs{\iota}_{n}$. For a matrix $\bs{A}$ which potentially has an increasing dimension, $\bs{\mathcal{O}}_p(1)$ is used to indicate that $||\bs{A}||_2 = \mathcal{O}_p(1)$ and similarly $\bs{\smallO}_p(1)$ signifies that $||\bs{A}||_2 = \smallO_p(1)$. Throughout, $c$ is used to denote some arbitrary positive constant, with indexation often indicating to which quantity the constant is associated; e.g. $c_{x}$ or $c_{\lambda}$, and `w.p.a.1' indicates `with probability approaching 1'.

\section{Estimation Approach}\label{motivation}
Introduced in \cite{bai_panel_2009}, the PC estimator is one of the foremost approaches taken to estimate models with interactive fixed effects, in situations where both $n$ and $T$ are large. This estimator is shown by the author to deliver consistent estimates of regression slope coefficients, and of rotational counterparts to the factors and the loadings, where the number of factors is known, and both $n$ and $T$ diverge. Further results have been provided by \cite{moon_linear_2015, moon_dynamic_nodate} who demonstrate that the estimator remains consistent with the number of factors unknown, but not underestimated, and also with the possible inclusion of predetermined regressors. These authors establish the asymptotic properties of the PC estimator and, in particular, document asymptotic biases that arise in the presence of cross-sectional and serial dependence, and due to inclusion of predetermined regressors. These biases originate from the incidental parameter problem, and although it is sometimes possible to mitigate their impact when both $n$ and $T$ are large, in situations where $T$ is small relative to $n$, or indeed fixed, they have proven to be more implacable, so much so that use of the PC estimator has been confined almost entirely to the large $n$, large $T$ setting. Yet, as is shown shortly, it is possible to resolve many of these issues by first transforming the model, and then going on to apply this estimator in the usual way. 

\subsection{Transformation}
It is useful to begin by re-writing the model in matrix form. Let the $n \times T$ matrix $\bs{Y} \coloneqq (\bs{y}_1,...,\bs{y}_T)$, $\bs{X}_{k}$ be the $n \times T$ matrix containing observations of the $k$-th covariate, the $T \times R^*$ matrix $\bs{F}^* \coloneqq (\bs{f}_1^*,...,\bs{f}_T^*)^\top$, and $\bs{S}(\alpha) \coloneqq \bs{I}_T - \alpha \bs{W}$, where $\bs{W}$ is a $T \times T$ shift matrix with zeros everywhere, except those elements directly above the main diagonal, which take a value of $1$. With this notation, the model can be written more succinctly as
\begin{align}
\bs{Y} \bs{S}(\alpha) 
= \sum^K_{k = 1} \beta_k \bs{X}_k + \bs{y}_0\bs{s}^\top(\alpha) + \bs{\Lambda}^*\bs{F}^{*\top} + \bs{\varepsilon}, \label{sys}
\end{align}
where 
\begin{align}
\bs{s}(\alpha)
&\coloneqq 
\begin{pmatrix}
\alpha & \bs{0}_{1 \times (T- 1)} 
\end{pmatrix}^\top.
\end{align}
In any dynamic panel model where $T$ is small, special care must be taken with the initial condition $\bs{y}_0\bs{s}^\top(\alpha)$ since this may itself be endogenous. In what follows the initial condition is treated as an additional parameter in the model and is absorbed into the factor term.\footnote{An alternative approach which involves dropping the first observation is discussed in Section \ref{estfac}.} As such, define $\bs{\Lambda} 
\coloneqq (\bs{y}_0,\bs{\Lambda}^*)$, $\bs{F}(\alpha) \coloneqq (\bs{s}(\alpha),\bs{F}^*)$, and $R \coloneqq R^* + 1$, whereby \eqref{sys} becomes
\begin{align}
\bs{Y} \bs{S}(\alpha) 
&= \sum^K_{k = 1} \beta_k \bs{X}_k +  \bs{\Lambda} \bs{F}^{\top} + \bs{\varepsilon}, \label{model}
\end{align}
with the dependence of $\bs{F}$ on $\alpha$ being suppressed. Now, define the $n \times TK$ matrix $\bs{\mathscr{X}} \coloneqq (\bs{X}_1,...,\bs{X}_K)$ which is assumed to have full column rank. Moreover, assume hereafter that $TK \leq n$.\footnote{Many of the results in this paper will carry over naturally to the small $n$, large $T$ setting by interchanging $n$ and $T$.} Consider the following group of transformations $\mathcal{{G}}$, where each element in this group is a bijective mapping from the sample space to itself:
\begin{align}
\mathcal{{G}}
\coloneqq \{ \bs{Q} \in \bs{\mathscr{O}}(n) : \bs{Q} \bs{\mathscr{X}} = \bs{\mathscr{X}} \}, \label{group}
\end{align}
with $\bs{\mathscr{O}}(n)$ being the group of $n \times n$ orthogonal matrices. This group $\mathcal{{G}}$ contains orthogonal transformations that preserve the column space of $\bs{\mathscr{X}}$. Take some $\bs{Q} \in \mathcal{{G}}$. This can be partitioned as
\begin{align}
\bs{Q} 
\eqqcolon 
\begin{pmatrix}
\bs{Q}_{\bs{\mathscr{X}}} & \bs{Q}_{\bs{\mathscr{X}}^\perp}
\end{pmatrix},
\end{align}
where $\bs{Q}_{\bs{\mathscr{X}}}$ is an $n \times TK$ matrix with orthonormal columns such that $\bs{Q}_{\bs{\mathscr{X}}}^\top \bs{Q}_{\bs{\mathscr{X}}} = \bs{I}_{TK}$ and $\bs{Q}_{\bs{\mathscr{X}}} \bs{Q}_{\bs{\mathscr{X}}}^\top = \bs{P}_{\bs{\mathscr{X}}}$, and, similarly, $\bs{Q}_{\bs{\mathscr{X}}^\perp}$ is an $n \times (n - TK)$ matrix with orthonormal columns such that $\bs{Q}_{\bs{\mathscr{X}}^\perp}^\top \bs{Q}_{\bs{\mathscr{X}^\perp}} = \bs{I}_{(n - TK)}$ and $\bs{Q}_{\bs{\mathscr{X}}^\perp} \bs{Q}_{\bs{\mathscr{X}}^\perp}^\top = \bs{M}_{\bs{\mathscr{X}}}$. Simply put, the matrix $\bs{Q}_{\bs{\mathscr{X}}}$ projects into the $TK$-dimensional space spanned by the columns of the matrix $\bs{\mathscr{X}}$, while $\bs{Q}_{\bs{\mathscr{X}^\perp}}$ on the other hand, projects into the space orthogonal to this. A simple way to construct 
$\bs{Q}_{\bs{\mathscr{X}}}$ would be as $\bs{\mathscr{X}}(\bs{\mathscr{X}}^\top\bs{\mathscr{X}})^{-\frac{1}{2}}$, and, with this in hand, the following transformed variables can be defined:
\begin{align}
\bs{\tilde{Y}}                  &\coloneqq \bs{Q}_{\bs{\mathscr{X}}}^\top \bs{Y}, \\	
\bs{\tilde{X}}_k                &\coloneqq \bs{Q}_{\bs{\mathscr{X}}}^\top \bs{X}_k, \\
\bs{\tilde{\Lambda}}            &\coloneqq \bs{Q}_{\bs{\mathscr{X}}}^\top \bs{\Lambda}, \\
\bs{\tilde{\varepsilon}}        &\coloneqq \bs{Q}_{\bs{\mathscr{X}}}^\top \bs{\varepsilon},
\end{align}
in which case premultiplying \eqref{model} by $\bs{Q}_{\bs{\mathscr{X}}}^\top$ yields the transformed model
\begin{align}
\bs{\tilde{Y}} \bs{S}(\alpha) 
&= \sum^K_{k = 1} \beta_k \bs{\tilde{X}}_k +  \bs{\tilde{\Lambda}} \bs{F}^{\top} + \bs{\tilde{\varepsilon}}. \label{model2}
\end{align}
Looking at \eqref{model2} there are three significant consequences of transforming the model through $\bs{Q}_{\bs{\mathscr{X}}}$ that need to be highlighted. First, the resultant matrices $\bs{\tilde{Y}}$, $\bs{\tilde{X}}_k$ and $\bs{\tilde{\Lambda}} \bs{F}^\top$ are of dimension $TK \times T$, since the entirety of the model has been transformed by $\bs{Q}_{\bs{\mathscr{X}}}$ and projected into the $TK$-dimensional subspace spanned by the columns of the covariates. Hence, the dimension of the factor term  $\bs{\tilde{\Lambda}} \bs{F}^\top$ will no longer depend on $n$, thereby reliving the model of incidental parameters as $n \rightarrow \infty$. Second, the transformation leads to no loss of information in the covariates since, by construction, transforming the model though $\bs{Q}_{\bs{\mathscr{X}}}$ preserves the columns space of $\bs{\mathscr{X}}$. Thirdly, since the covariates used in the construction of $\bs{Q}_{\bs{\mathscr{X}}}$ are strictly exogenous, under quite general conditions, including broad cross-sectional and serial dependence, the transformation renders the error term asymptotically negligible (in a precise scene) so long as $T/n \rightarrow 0$. Indeed, it is this final property in particular that will prove key to estimation of \eqref{model2} by principal components.

\subsection{Estimation by Principle Components}\label{EstPC}
The intuition underlying the PC estimator is that, given the factors and the loadings, the coefficients can be estimated by least squares, and, similarly, given $\bs{\theta}$, estimating the factors and loadings is a standard principal component problem. Where $T$ is small relative to $n$, it is this latter step that proves to be challenging; in particular estimating the $n$-dimensional factor loadings. For this reason it is useful to consider the factor term in isolation in order to demonstrate the key differences that lie between PC estimation of the original model, and of its transformed counterpart.  

Assume that $\bs{\theta}$ is observed and define $\bs{\dot{Y}} \coloneqq \bs{Y} \bs{S}(\alpha) - \sum^K_{k=1}\beta_{k} \bs{X}_k = \bs{\Lambda}\bs{F}^\top + \bs{\varepsilon}$ which has a pure factor structure. Let $\bs{\check{\Lambda}}$ and $\bs{\check{F}}$ be $n \times R$ and $T \times R$ matrices, respectively, which satisfy $\bs{\check{\Lambda}} \bs{\check{F}}^\top = \bs{{\Lambda}} \bs{{F}}^\top$,  $\frac{1}{n}\bs{\check{\Lambda}}^\top \bs{\check{\Lambda}} = \bs{I}_R$ and $\bs{\check{F}}^\top \bs{\check{F}}$ being diagonal.\footnote{It is straightforward to see that such matrices exist. For example, by the singular value decomposition, decompose $\bs{\Lambda}\bs{F}^\top = \bs{U}\bs{S}\bs{V}^\top$. Let $\bs{\check{\Lambda}}$ be the $R$ columns of $\sqrt{n}\bs{U}$ associated with the nonzero singular values and $\bs{\check{F}}^\top$ be the corresponding $R$ rows of $\bs{S}\bs{V}^\top/\sqrt{n}$. As the columns of $\bs{U}$ and $\bs{V}$ are orthogonal, and $\bs{S}$ is diagonal, it follows that $\bs{\check{\Lambda}}^\top\bs{\check{\Lambda}}/n = \bs{I}_R$, $\bs{\check{F}}^\top\bs{\check{F}}$ is diagonal and $\bs{\check{\Lambda}}\bs{\check{F}}^\top = \bs{\Lambda}\bs{F}^\top$.} Consider the problem of trying to estimate the loadings from the variance of $\bs{\dot{Y}}$.
\begin{align}
\frac{1}{nT} \bs{\dot{Y}} \bs{\dot{Y}}^\top \bs{\check{\Lambda}} 
&= \frac{1}{nT} (\bs{\Lambda}\bs{F}^\top + \bs{\varepsilon}) (\bs{\Lambda}\bs{F}^\top + \bs{\varepsilon})^\top \bs{\check{\Lambda}}  \notag \\
&= \frac{1}{nT} \bs{\Lambda}\bs{F}^\top \bs{F} \bs{\Lambda}^\top  \bs{\check{\Lambda}}  
+ \frac{1}{nT} \bs{\Lambda}\bs{F}^\top \bs{\varepsilon}^\top \bs{\check{\Lambda}}  
+ \frac{1}{nT}\bs{\varepsilon}\bs{F}\bs{\Lambda}^\top \bs{\check{\Lambda}}  
+ \frac{1}{nT} \bs{\varepsilon}  \bs{\varepsilon}^\top \bs{\check{\Lambda}}.  \label{eq8} 
\end{align}
With suitable conditions on the errors, the factors, and the loadings, as $n \rightarrow \infty$,
\begin{align}
\frac{1}{nT} \bs{\dot{Y}} \bs{\dot{Y}}^\top \bs{\check{\Lambda}} 
&= \frac{1}{nT} \bs{\check{\Lambda}}\bs{\check{F}}^\top \bs{\check{F}}  + \frac{1}{nT} \bs{\varepsilon}  \bs{\varepsilon}^\top \bs{\check{\Lambda}}  + \bs{\smallO}_p \left( 1 \right). \label{eq9}
\end{align}
Given that $\frac{1}{n}\bs{\check{\Lambda}}^\top \bs{\check{\Lambda}} = \bs{I}_R$ and $\bs{\check{F}}^\top \bs{\check{F}}$ is diagonal, then, without the term $\frac{1}{nT} \bs{\varepsilon}  \bs{\varepsilon}^\top \bs{\check{\Lambda}}$, $\bs{\check{\Lambda}}$ would be an eigenvector of $\frac{1}{nT} \bs{\dot{Y}}\bs{\dot{Y}}^\top$ asymptotically, and would provide a rotational counterpart to the factor loadings $\bs{\Lambda}$. Where both $n$ and $T$ are large, several authors have shown that, in spite of this distortionary term, estimating the loadings in the manner above is still possible in certain circumstances. For example, under the condition $||\bs{\varepsilon}||_2 =\mathcal{O}_p(\sqrt{\max\{n,T\}})$ employed in \cite{moon_linear_2015}, dependence in the error term is sufficiently limited that $\frac{1}{nT} \bs{\varepsilon}  \bs{\varepsilon}^\top \bs{\check{\Lambda}} = \bs{\smallO}_p \left( 1 \right)$ as $n,T \rightarrow \infty$. Alternatively, where $\frac{1}{nT} \bs{\varepsilon} \bs{\varepsilon}^\top  \xrightarrow{p} \bs{\Sigma}_{\bs{\varepsilon}}$, it may be possible to jointly estimate $\bs{\Sigma}_{\bs{\varepsilon}}$, and then a rotation of $\bs{\Lambda}$ as an eigenvector of $\frac{1}{nT} \bs{\dot{Y}} \bs{\dot{Y}}^\top - \bs{\Sigma}_{\bs{\varepsilon}}$. Nonetheless, in either case it is only under the most exceptional of circumstances that the distortions caused by $\frac{1}{nT} \bs{\varepsilon} \bs{\varepsilon}^\top \bs{\check{\Lambda}}$ do not affect the estimation of the parameter $\bs{\theta}$, and, moreover, neither case generally applies to the situation where $T$ is fixed.

Now consider, on the other hand, the transformed model. Let $\bs{\check{\tilde{\Lambda}}}$ denote an analogue of $\bs{\check{{\Lambda}}}$. With $\frac{1}{nT} ||\bs{\tilde{\varepsilon}} \bs{F} \bs{{\tilde{\Lambda}}}^\top \bs{\check{\tilde{\Lambda}}} ||_2 = {\smallO}_p \left( 1 \right)$, one arrives at a similar expression to \eqref{eq9}
\begin{align}
\frac{1}{nT} \bs{\dot{\tilde{Y}}} \bs{\dot{\tilde{Y}}}^\top \bs{\check{\tilde{\Lambda}}}
=&\ \frac{1}{nT} \bs{\check{\tilde{\Lambda}}} \bs{\check{F}}^\top \bs{\check{F}} 
+ \frac{1}{nT} \bs{\tilde{\varepsilon}} \bs{\tilde{\varepsilon}}^\top  \bs{\check{\tilde{\Lambda}}} + \bs{\smallO}_p \left( 1 \right). \label{eq13}
\end{align}
Yet now, since the regressors used to construct $\bs{Q}_{\bs{\mathscr{X}}}$ are strictly exogenous, under quite general conditions $ \frac{1}{nT}|| \bs{\tilde{\varepsilon}} \bs{\tilde{\varepsilon}}^\top  ||_2 = \frac{1}{nT}|| \bs{\varepsilon}^\top \bs{P}_{\bs{\mathscr{X}}}  \bs{\varepsilon}||_2 = \smallO_p(1)$, even with fixed $T$. As a consequence, asymptotically, $\bs{\check{\tilde{\Lambda}}}$ will be an eigenvector of $\frac{1}{nT} \bs{\dot{\tilde{Y}}}  \bs{\dot{\tilde{Y}}}^\top$ and thus it is possible to estimate the space spanned by $\tilde{\boldsymbol{\Lambda}}$ with fixed $T$, where this was not possible for $\bs{\Lambda}$. This, heuristically, is why applying the PC estimator to the transformed model is able to deliver consistent estimates with $T$ fixed. 

\subsection{Objective Function}
Following \cite{moon_linear_2015}, the transformed model \eqref{model2} can be estimated by minimising the following least squares objective function:
\begin{align}
\mathcal{Q}(\bs{\theta},\bs{\tilde{\Lambda}},\bs{F})
&\coloneqq \frac{1}{nT}\text{tr}\left(\left(\bs{\tilde{Y}} \bs{S}(\alpha) - \sum^{K}_{k=1} \beta_k \bs{\tilde{X}}_k  - \bs{\tilde{\Lambda}} \bs{F}^\top\right)^\top \left(\bs{\tilde{Y}} \bs{S}(\alpha) - \sum^{K}_{k=1} \beta_k \bs{\tilde{X}}_k - \bs{\tilde{\Lambda}} \bs{F}^\top\right)\right).
\label{eq6}
\end{align}\normalsize
Both the factors and the transformed loadings can be concentrated out of \eqref{eq6}, in which case one arrives at an objective function involving $\bs{\theta}$ alone,
\begin{align}
\mathcal{Q}(\bs{\theta}) 
\coloneqq \frac{1}{nT} \sum^T_{r=R+1} \mu_{r} \left(  \left(\bs{\tilde{Y}} \bs{S}(\alpha) - \sum^{K}_{k=1} \beta_k \bs{\tilde{X}}_k \right)^\top  \left(\bs{\tilde{Y}} \bs{S}(\alpha) - \sum^{K}_{k=1} \beta_k \bs{\tilde{X}}_k \right) \right), \label{eq7}
\end{align}
that is, the profile objective function now involves the sum of the $(T-R)$ smallest eigenvalues of the right-hand matrix.\footnote{See Appendix A.1 for details.} Using this, the estimator $\bs{\hat{\theta}}$ can then be defined as 
\begin{align}
\bs{\hat{\theta}} \coloneqq \argmin_{\bs{\theta} \in {\Theta}}  \mathcal{Q}(\bs{\theta}). \label{defest}
\end{align}

\begin{remark}
	Since the PC estimator draws on the factor model literature in several important ways, it is unsurprising that, in order to simultaneously estimate the $n$-dimensional factor loadings, and $T$-dimensional factors, it is usually necessary that both $n$ and $T$ diverge. This can be related to results on pure factor models, where only exceptionally will principal component methods produce consistent estimates of both factors and loadings with fixed $T$; see, for instance, Theorem 4 of \cite{baifac}.
\end{remark}

\begin{remark}
	Reducing the dimension of the factor term may relieve the model of incidental parameters in the cross-section, but the effect of these parameters does not disappear entirely. Their effect is still present through $\bs{\tilde{\Lambda}}$, the part of the factor loadings that remains, which manifests itself as an additional incidental parameter in the time dimension. 
\end{remark}

\begin{remark}
	This paper focuses on the case where, with the exception of lagged outcomes, the regressors are strictly exogenous, as in \cite{bai_panel_2009} and \cite{moon_linear_2015}. If instead some of the covariates $\bs{X}_k$ are endogenous, but valid instruments for these are available, then those instruments can substitute for $\bs{X}_k$ in the construction of $\bs{\mathscr{X}}$.   	
\end{remark}

\begin{remark}
	When estimating the original model, the least squares objective function can be interpreted as the negative of a quasi-maximum likelihood function that uses the standard normal distribution. The objective function \eqref{eq6} can then be interpreted as a marginal quasi-likelihood which uses only a part of this. 
\end{remark}

\section{Consistency}\label{consub}
This section studies the asymptotic properties of the estimator defined in \eqref{defest}, beginning by deriving a general consistency result. As in \cite{moon_linear_2015}, throughout the following, both $\bs{\Lambda}$ and $\bs{F}$ are treated as fixed parameters in estimation and the superscript $0$ is now introduced to distinguish true parameter values. Moreover, let $\bs{S} \coloneqq \bs{S}(\alpha^0)$, $\bs{G} \coloneqq \bs{S}^{-1}\bs{W}$, and $\mathcal{C}$ denote $\sigma(\bs{X}_1,...,\bs{X}_K)$, that is, the sub-algebra generated by the exogenous covariates. The following assumptions are made.
\begin{AssumptionMD}[Model]\label{AMD} \color{white}.\color{black}\
	\begin{enumerate}[label=\textup{(\roman*)}]
		\item The parameter vector $\bs{\theta}^0$ lies in the interior of ${\Theta}$, where ${\Theta}$ is a compact subset of $\mathbb{R}^{K+1}$ in which $|\alpha|<1$. \label{AMDi}
		\item The elements of the matrices $\bs{X}_1,...,\bs{X}_K$, $\bs{\Lambda}^0$ and $\bs{F}^0$ have uniformly bounded fourth moments. \label{AMDii} 
	\end{enumerate}
\end{AssumptionMD}
Assumption \hyperref[AMDi]{{MD(i)}} assumes that the dynamic process is stationary, which allows $\bs{y}_t$ to be expanded as an infinite series by recursive substitution. Assumption \hyperref[AMDii]{{MD(ii)}} imposes standard conditions on the moments of the covariates, the factors, and the loadings. 
\begin{AssumptionER}[Error]\label{ADE}  \color{white}.\color{black}\
	\begin{enumerate}[label=\textup{(\roman*)}]
		\item $\mathbb{E}[\varepsilon_{it}|\mathcal{C}] = 0$ for $i = 1,...,n$, $t= 1,...,T$. \label{ERi} 
		\item Let $\sigma^2_{ij,t\tau} = \mathbb{E}[{\varepsilon}_{it}\varepsilon_{j\tau}|\mathcal{C}]$. Then $|\sigma^2_{ij,t\tau}| < C$ uniformly for all $i,j,t,\tau$, and the error term is weakly conditionally cross-sectionally and serially dependent, that is, $\sum_{i \neq j} |\sigma^2_{ij,t\tau}| \leq C$ uniformly for all $j,t,\tau$, and $\sum_{t \neq \tau} |\sigma^2_{ij,t\tau}| \leq C$ uniformly for all $i,j,\tau$. \label{ERii}
	\end{enumerate}
\end{AssumptionER}
Assumption \hyperref[ERi]{ER(i)} imposes strict exogeneity of the regressors as in \cite{bai_panel_2009} and \cite{moon_linear_2015}. Assumption \hyperref[ADE]{ER(ii)} limits the degree of dependence between the errors in the cross-section and across time, while allowing for heteroskedasticity in both dimensions of the panel. Different notions of dependence appear throughout the panel literature, and this can be modelled in several ways. Assumption \hyperref[ERii]{ER(ii)} is quite general in this regard.  

It is important to point out that the least squares objective function given in \eqref{eq7} implicitly uses the reduced form of the dynamic process to generate an internal instrument for the autoregressive parameter. To see this, notice that $\bs{S}^{-1}(\alpha) = \bs{I}_T + \alpha \bs{G}(\alpha)$. Substituting this into the reduced form then yields
\begin{align}
	\bs{\tilde{Y}} 
	&= \left( \sum^K_{k=1} \beta_k \bs{\tilde{X}}_k + \bs{\tilde{\Lambda}}\bs{F}^\top + \bs{\tilde{\varepsilon}} \right) \bs{S}^{-1}(\alpha) \notag \\
	&= \alpha \left( \sum^K_{k=1} \beta_k \bs{\tilde{X}}_k \bs{G}(\alpha)  \right)  +  \sum^K_{k=1} \beta_k \bs{\tilde{X}}_k + (\bs{\tilde{\Lambda}}\bs{F}^\top + \bs{\tilde{\varepsilon}}) \bs{S}^{-1}(\alpha).
\end{align}
In this way the role that $\sum^K_{k=1} \beta_k \bs{\tilde{X}}_k \bs{G}(\alpha)$ plays as an instrument for $\alpha$ is clear. Going forward it is useful to collect this instrument and the other exogenous covariates into a single matrix of regressors. Therefore let $\bs{\tilde{Z}}_1 \coloneqq \sum^K_{k=1} \beta_k \bs{\tilde{X}}_k \bs{G}(\alpha)$, $\bs{\tilde{Z}}_{k+1} \coloneqq \bs{\tilde{X}}_k$ for $k = 1,..,K$, $\bs{\delta} \cdot \bs{\tilde{{Z}}} \coloneqq \sum^{K+1}_{\kappa=1} \delta_k \bs{\tilde{{Z}}}_\kappa$ and define $\bs{\tilde{\mathcal{Z}}} \coloneqq  (\text{vec}(\bs{\tilde{Z}}_1),...,\text{vec}(\bs{\tilde{Z}}_{K+1})) \in \mathbb{R}^{KT^2 \times (K+1)}$.
\begin{AssumptionCS}[Consistency]\label{ACS}\color{white}.\color{black}\
	\begin{enumerate}[label=\textup{(\roman*)}]
		\item $R \geq R^0 \coloneqq \textup{rank}(\bs{\tilde{\Lambda}}^0\bs{F}^{0^\top})$. \label{ACSi}
		\item $\min_{\bs{\delta} \in \mathbb{R}^{K+1}: ||\bs{\delta}||_2 = 1} \sum^T_{r = R + R^0 + 1} \mu_r \left( \frac{1}{nT} (\bs{\delta} \cdot \bs{\tilde{{Z}}})^\top (\bs{\delta} \cdot \bs{\tilde{{Z}}}) \right) \geq b > 0$. \label{ACSii}
	\end{enumerate} 
\end{AssumptionCS}
Assumption \hyperref[ACSi]{{CS(i)}} allows for the true number of factors ${R}^0$ to be unknown as long as the number of factors used in estimation $R$ is no less than $R^0$. Notice also that this condition concerns the rank of $\bs{\tilde{\Lambda}}^0\bs{F}^{0^\top}$ and not of $\bs{{\Lambda}}^0\bs{F}^{0^\top}$, that is, $R^0$ is the number of factors correlated with the covariates. Assumption \hyperref[ACSii]{{CS(ii)}} is a multicollinearity condition and can intuitively be understood by realising that it implies $\inf_{\bs{\tilde{\Lambda}} \in \mathbb{R}^{TK \times R^0}, \bs{F} \in \mathbb{R}^{T \times R}} \mu_{K+1}(\bs{\mathcal{\tilde{Z}}}^\top (\bs{M}_{\bs{F}} \otimes \bs{M}_{\bs{\tilde{\Lambda}}}) \bs{\mathcal{\tilde{Z}}})$ is bounded away from zero. This, therefore, asserts that the data matrix $\bs{\tilde{\mathcal{Z}}}^\top \bs{\tilde{\mathcal{Z}}}$ retains a sufficient level of variation, after having been projected orthogonal to arbitrary $R \times T$ factors and $R^0 \times TK$ loadings. 
\begin{proposition}[Consistency -- General]\label{CS} 
	Under Assumptions \hyperref[AMD]{{MD}}, \hyperref[ADE]{{ER}} and \hyperref[ACS]{{CS}}, 
	\begin{align}
		||\bs{\hat{\theta}} - \bs{\theta}^0||_2 = \mathcal{O}_p \left( \sqrt{\frac{T}{n}} \right). 
	\end{align} 
\end{proposition}
Proposition \ref{CS} demonstrates that as $T/n \rightarrow 0$ the estimator is consistent. Moreover, where $T$ is fixed, it is in fact $\sqrt{n}$ consistent. This result is obtained under quite general dependence in the error, and as long as the number of factors used in estimation is no less than the true number. Notice also that no assumptions have been made regarding the factors and the loadings other than bounded fourth moments; for instance, these may be strong, weak, or non-existent. 

Proposition \ref{CS} can be compared directly to Theorem 4.1 in \cite{moon_linear_2015}, which, under similar terms, provides a consistency result for the PC estimator applied to the original model. Their result establishes that
\begin{align}
||\bs{\theta}^0 - \bs{\hat{\theta}}||_2 
= \mathcal{O}_p\left( \frac{1}{\sqrt{\min\{n,T\}}} \right), 
\end{align}
with this rate being determined largely by the condition $||\bs{\varepsilon}||_2 = \mathcal{O}_p(\sqrt{\max\{n,T\}})$ (Assumption SN(ii)), under which\footnote{Moreover, \eqref{asr} also proves to be important for the asymptotic expansion of the objective function; see Section \ref{asyexs}.} 
\begin{align}
	\frac{||\bs{\varepsilon}||_2}{\sqrt{nT}} = \mathcal{O}_p\left( \frac{1}{\sqrt{\min\{n,T\}}} \right). \label{asr}
\end{align}
In similar fashion, the rate obtained in Proposition \ref{CS} can be attributed to the quantity $||\bs{\tilde{\varepsilon}}||_F$ which plays an analogous role in this paper. Under Assumption \hyperref[ERi]{ER} this can be shown to satisfy 
\begin{align}
\frac{||\bs{\tilde{\varepsilon}}||_F}{\sqrt{nT}} 
= \mathcal{O}_p\left(\sqrt{\frac{T}{n}}\right).
\end{align}
Recalling the discussion in Section \ref{EstPC}, it is worth stressing again that while the difference between the quantities $\bs{\varepsilon}$ and $\bs{\tilde{\varepsilon}}$ may appear superficial, it is in fact of the utmost significance. For example, consider the textbook assumption of identically and independently distributed conditionally homoskedastic errors; i.e. $\mathbb{E}[\varepsilon_{it}\varepsilon_{j\tau}|\mathcal{C}] = \sigma^2$ for $i = j$, $t = \tau$ and is zero otherwise. In this case,
\begin{align}
	\mathbb{E} [||\bs{\tilde{\varepsilon}}||_F^2] 
	= \mathbb{E} [||\bs{Q}_{\bs{\mathscr{X}}}^\top\bs{{\varepsilon}}||_F^2] 
	&= \mathbb{E} \left[ \sum^{TK}_{t=1} \sum^T_{\tau=1} \sum^n_{j=1} \sum^n_{i=1} \mathbb{E}[ ({Q}_{{\mathscr{X}}})_{it} ({Q}_{{\mathscr{X}}})_{jt} \varepsilon_{i\tau} \varepsilon_{j\tau}  | \mathcal{C}]  \right] \notag \\
	&= \sigma^2  \sum^{TK}_{t=1} \sum^T_{\tau=1}  \sum^n_{i=1} \mathbb{E}[ ({Q}_{{\mathscr{X}}})_{it}^2] \notag \\
	&= \sigma^2 T  \mathbb{E} \left[ ||\bs{Q}_{\bs{\mathscr{X}}}||_F^2  \right] \notag \\
	&= \sigma^2T^2K 
	= \mathcal{O}(T^2),
\end{align}
from which it then follows by Markov's inequality that $||\bs{\tilde{\varepsilon}}||_F = \mathcal{O}_p(T)$, and so
\begin{align}
	\frac{||\bs{\tilde{\varepsilon}}||_F}{\sqrt{nT}} 
	= \mathcal{O}_p\left(\sqrt{\frac{T}{n}}\right) 
	= \smallO_p(1),
\end{align}
as $T/n \rightarrow 0$. By comparison, 
\begin{align}
	\frac{||\bs{\varepsilon}||_2}{\sqrt{nT}} 
	\geq \frac{1}{\sqrt{nT}} \frac{1}{\sqrt{\min\{n,T\}}}||\bs{\varepsilon}||_F \xrightarrow{p} \frac{\sigma}{\sqrt{\min\{n,T\}}},
\end{align}
using $\frac{1}{\sqrt{\text{rank}(\bs{A})}}||\bs{A}||_F \leq ||\bs{A}||_2$ and noting that, in this particular example,
\begin{align}
	\frac{||\bs{\varepsilon}||_F}{\sqrt{nT}}  = \frac{1}{\sqrt{nT}}  \sqrt{\sum^T_{t=1}\sum^n_{i=1}\varepsilon_{it}^2} \xrightarrow{p} \sigma. 
\end{align}
Therefore, even in this simple case, $\frac{||\bs{\varepsilon}||_2}{\sqrt{nT}}$ cannot be $\smallO_p(1)$ with $T$ fixed, as long as $\sigma$ is bounded from below by a constant. 

\begin{remark}
	\cite{bai_panel_2009} also obtains an initial consistency result under weaker conditions on the errors than $||\bs{\varepsilon}||_2 = \mathcal{O}_p(\sqrt{\min\{n,T\}})$. However, this result is obtained assuming that $R = R^0$, and the factors and loadings are independent of the errors. Neither of these are assumed in Proposition \ref{CS}.
\end{remark}


\section{Asymptotic Expansion}\label{asyexs}
Typically the asymptotic distribution of an extremum estimator is obtained by expanding the objective function locally around the true parameter value. It is, however, difficult to obtain an expansion of the objective function \eqref{eq7} since this involves a summation over a certain number of eigenvalues of a matrix. Following \cite{bai_panel_2009}, an alternative approach would be to proceed from the first order conditions of the optimisation problem, to avoid dealing with the fully concentrated objective function. Yet \cite{moon_linear_2015} show that it is possible to analyse this objective function directly, by utilising perturbation theory for linear operators to derive an expansion for the perturbed eigenvalues of $\bs{F}^0{\bs{\tilde{\Lambda}}^{0}}^\top\bs{\tilde{\Lambda}}^0\bs{F}^{0^\top}/nT$. Key to this approach is demonstrating that the perturbation is asymptotically small, which in this case follows from Proposition \ref{CS}, whereby $|{\theta}^0_{\kappa} - {\hat{\theta}}_{\kappa}|$ is small, and from assuming that the `perturbation' stemming from the error term, $\frac{||\bs{\tilde{\varepsilon}}||_2}{\sqrt{nT}}$, diminishes asymptotically. In light of the discussion in the previous section, the significance of transforming the errors is again highlighted as expansion of the objective function remains valid only so long as $\frac{||\bs{\tilde{\varepsilon}}||_2}{\sqrt{nT}}$ is asymptotically small. Since $||\bs{\tilde{\varepsilon}}||_2 \leq ||\bs{{\varepsilon}}||_2$, $\frac{||\bs{\tilde{\varepsilon}}||_2}{\sqrt{nT}}$ will be asymptotically small in situations where this will not be  true of $\frac{||\bs{{\varepsilon}}||_2}{\sqrt{nT}}$.\footnote{The inequality $||\bs{\tilde{\varepsilon}}||_2 \leq ||\bs{\varepsilon}||_2$ is obtained by the 
submultiplicity of the spectral norm and noting that $||\bs{Q}_{\bs{\mathscr{X}}}||_2 = 1$.}
\begin{AssumptionAE}[Asymptotic Expansion]\label{AAE} \color{white}.\color{black}\
	\begin{enumerate}[label=\textup{(\roman*)}]
		\item $R = R^0$. \label{AEi}
		\item $\frac{1}{n} \bs{\tilde{\Lambda}}^{0\top} \bs{\tilde{\Lambda}}^0 = \frac{1}{n} \bs{\Lambda}^{0\top} \bs{P}_{\bs{\mathscr{X}}} \bs{\Lambda}^0 \xrightarrow{p} \bs{\Sigma}_{ \bs{\tilde{\Lambda}}^0 }$ as $n \rightarrow \infty$, with $\mu_{R^0}(\bs{\Sigma}_{ \bs{\tilde{\Lambda}}^0}) > 0$ and $\mu_{1}(\bs{\Sigma}_{ \bs{\tilde{\Lambda}}^0}) < \infty$. \label{AEii}
		\item $\frac{1}{T} \bs{F}^{0\top} \bs{F}^0 \xrightarrow{p} \bs{\Sigma}_{\bs{F}^0} > 0$ as $T \rightarrow \infty$, with $\mu_{R^0}(\bs{\Sigma}_{\bs{F}^0}) > 0$ and $\mu_{1}(\bs{\Sigma}_{\bs{F}^0}) < \infty$. \label{AEiii}
	\end{enumerate}
\end{AssumptionAE}
In the absence of dynamics, \cite{moon_linear_2015} show that, under certain conditions, the asymptotic distribution of the PC estimator is unaffected by overstatement of the number of factors. Though it might also be expected that a similar result could be obtained in the present case, the asymptotic distribution is derived under the assumption that the number of factors is correctly specified; that is $R  = R^0$ as in Assumption \hyperref[AEi]{{AE(i)}}. A method to detect the true number of factors is discussed in Section \ref{estfac}. Assumptions \hyperref[AEii]{{AE(ii)}} and \hyperref[AEiii]{{AE(iii)}} assume the factors and the transformed factor loadings are strong and both have a nonnegligible impact on the variance of the term $\bs{\tilde{\Lambda}}^0 \bs{F}^{0^\top} + \bs{\tilde{\varepsilon}}$. 
\begin{proposition}[Asymptotic Expansion]\label{AE} 
	Under Assumptions \hyperref[AMD]{{MD}}, \hyperref[ADE]{{ER}} and \hyperref[AAE]{{AE}}, if $||\bs{\theta}^0 - \bs{{\theta}}||_2 = \smallO_p(1)$, then, as $T^2/n \rightarrow 0$,
	\begin{align}
		\mathcal{Q}(\bs{\theta})
		=&\ \mathcal{Q}(\bs{\theta}^0) 
		- \frac{2}{\sqrt{nT}}(\bs{\theta}-\bs{\theta}^0)^\top \bs{d} + (\bs{\theta}-\bs{\theta}^0)^\top \bs{D} (\bs{\theta}-\bs{\theta}^0) +  \bs{r}(\bs{\theta}),
	\end{align} 
	where $\bs{d} \coloneqq \bs{c} + \bs{b}^{(1)} + \bs{b}^{(2)} + \bs{b}^{(3)} + \bs{b}^{(4)}$, the elements of these vectors and matrices are given by
	\begin{align}
		{D}_{\kappa\kappa'}
		&\coloneqq \frac{1}{nT} \textup{tr}(\bs{\tilde{Z}}_\kappa \bs{M}_{\bs{F}^0}\bs{\tilde{Z}}_{\kappa'}^{\top} \bs{M}_{\bs{\tilde{\Lambda}}^0} ),\label{ddef}  \\
		c_{\kappa} 
		&\coloneqq \frac{1}{nT} \textup{tr}(\bs{\tilde{Z}}_\kappa \bs{M}_{\bs{F}^0} \bs{\tilde{\varepsilon}}^\top \bs{M}_{\bs{\tilde{\Lambda}}^0} ),  \\
		b^{(1)}_{1} 
		&\coloneqq
		\frac{1}{\sqrt{nT}} \textup{tr}\left(\bs{M}_{\bs{F}^0} \bs{G} \bs{M}_{\bs{F}^0} \bs{\tilde{\varepsilon}}^{\top} \bs{M}_{\bs{\tilde{\Lambda}}^0} \bs{\tilde{\varepsilon}}  \right) , \\		
		b^{(2)}_{\kappa}
		&\coloneqq -\frac{1}{\sqrt{nT}}  \textup{tr}\left(\bs{M}_{\bs{F}^0} \bs{\tilde{\varepsilon}}^{\top}  \bs{M}_{\bs{\tilde{\Lambda}}^0}  \bs{{\tilde{Z}}}_\kappa \bs{F}^0 (\bs{F}^{0\top} \bs{F}^0)^{-1} (\bs{\tilde{\Lambda}}^{0\top} \bs{\tilde{\Lambda}}^0)^{-1} \bs{\tilde{\Lambda}}^{0\top}  \bs{\tilde{\varepsilon}} \right), \\
		b^{(3)}_{\kappa}
		&\coloneqq- \frac{1}{\sqrt{nT}} \textup{tr}\left(\bs{M}_{\bs{F}^0} \bs{\tilde{Z}}_\kappa^{\top}  \bs{M}_{\bs{\tilde{\Lambda}}^0} \bs{\tilde{\varepsilon}}\bs{F}^0 (\bs{F}^{0\top} \bs{F}^0)^{-1} (\bs{\tilde{\Lambda}}^{0\top} \bs{\tilde{\Lambda}}^0)^{-1} \bs{\tilde{\Lambda}}^{0\top} \bs{\tilde{\varepsilon}} \right),  \\
		b^{(4)}_{\kappa}
		&\coloneqq -\frac{1}{\sqrt{nT}}  \textup{tr}\left(\bs{M}_{\bs{F}^0} \bs{\tilde{\varepsilon}}^{\top} \bs{M}_{\bs{\tilde{\Lambda}}^0} \bs{\tilde{\varepsilon}} \bs{F}^0 (\bs{F}^{0\top} \bs{F}^0)^{-1} (\bs{\tilde{\Lambda}}^{0\top} \bs{\tilde{\Lambda}}^0)^{-1} \bs{\tilde{\Lambda}}^{0\top}  \bs{\tilde{Z}}_{\kappa}   \right),
	\end{align}
	and $b^{(1)}_{\kappa} \coloneqq 0$ for $\kappa = 2,...{K+1}$. Moreover, the order of the term $\bs{r}(\bs{\theta})$ is $\bs{\smallO}_p \left( \frac{ (1 + \sqrt{nT}||\bs{\theta}^0 - \bs{{\theta}}||_2)^2}{nT} \right)$.
\end{proposition}


As will be seen in the subsequent section, the term $\bs{c}$ plays a central role in determining the asymptotic distribution of the estimator. Term $\bs{b}^{(1)}$ arises due to the presence of a lagged outcome. When applying the PC estimator to the original model, an equivalent term arises and is the source of a bias, as described in \cite{moon_dynamic_nodate}. Terms $\bs{b}^{(2)}, \bs{b}^{(3)}$ and $\bs{b}^{(4)}$ appear due to cross-sectional and serial dependence in the error term, and, again, have corresponding terms described in both \cite{bai_panel_2009} and \cite{moon_linear_2015,moon_dynamic_nodate} which give rise to additional asymptotic biases. Under Assumptions \hyperref[AMD]{{MD}}, \hyperref[ADE]{{ER}} and \hyperref[AAE]{{AE}}, it can be established that $\bs{b}^{(1)}, \bs{b}^{(2)},\bs{b}^{(3)}$ and $\bs{b}^{(4)}$ are $\bs{\mathcal{O}}_p(T^{1/5}/\sqrt{n})$ which suggests that the estimator is asymptotically unbiased where $T^3/n \rightarrow 0$. This is of course trivially satisfied where $T$ is fixed. Using this, and the expression given in Proposition \ref{AE}, the following result can be obtained. 
\begin{proposition}[$\sqrt{nT}$ Consistency]\label{sqnTCS}	
	Under Assumptions \hyperref[AMD]{{MD}}, \hyperref[ADE]{{ER}}, \hyperref[ACS]{{CS}} and \hyperref[AAE]{{AE}}, and assuming that $||\bs{c}||_2 = \mathcal{O}_p(1)$, then, as $T^3/n \rightarrow 0$, 	
	\begin{align}
		\sqrt{nT} (\bs{\hat{\theta}} - \bs{\theta}^0) 
		=  \bs{\mathcal{O}}_p \left( 1 \right).
	\end{align} 
\end{proposition}
Proposition \ref{sqnTCS} verifies the $\sqrt{n}$ consistency of the estimator in the event that $T$ is fixed, and also that with the number of factors known, and $n$ increasing sufficiently fast relative to $T$, $\sqrt{nT}$ consistency of the estimator can be obtained as both $n$ and $T \rightarrow \infty$. The origin of the condition $T^3/n \rightarrow 0$ is explained more fully in the next section where it is shown to arise from the inclusion of a dynamic regressor and that, under stronger conditions, it is possible to reduce this to $T/n \rightarrow 0$. 

\section{Asymptotic Distribution}\label{asydist}
This section studies the asymptotic distribution of the estimator, culminating in the central result of this paper which establishes that, under more restrictive conditions, with $T$ fixed, the estimator is asymptotically unbiased despite the presence of cross-sectional dependence, serial dependence, and with the inclusion of dynamic regressors. However, in order to really appreciate the impact that transforming the model through $\bs{Q}_{\bs{\mathscr{X}}}$ has, this section first presents a more general result which is derived under the assumption that $T/n \rightarrow c$ with $c \in [0,K^{-1}]$. This is useful to paint a more complete picture of the asymptotic properties of the estimator, and leads to a greater appreciation of the fundamental effects of transforming the model. To begin, some additional assumptions are introduced that are utilised in both cases. 

\begin{AssumptionER*}[Error]\label{ER*}
The errors are generated as $\bs{\varepsilon} = \bs{\Sigma}^{\frac{1}{2}}_{n}\bs{U}\bs{\Sigma}^{\frac{1}{2}}_{T}$, where $\bs{\Sigma}_{n}^{\frac{1}{2}}$ and $\bs{\Sigma}_{T}^{\frac{1}{2}}$ are symmetric matrices of dimension $n \times n$ and $T \times T$, respectively, both of which are uniformly bounded in absolute row and column sums, and have eigenvalues uniformly bounded from below by a positive constant. $\bs{U}$ is an $n \times T$ matrix, with elements $u_{it}$ which are independent of the exogenous covariates, the factors and the loadings, and identically and independently distributed across $i$ and $t$, with $\mathbb{E}[u_{it}] = 0$, $\mathbb{E}[u_{it}^2] = 1$ and $\mathbb{E}[u_{it}^4] \leq C$. \label{ER(i)}
\end{AssumptionER*}
Assumption \hyperref[ER*]{ER*} still allows for heteroskedasticity in both dimensions of the panel, but serves to place further limits on possible dependence in the errors across the cross-section and between time periods. Moreover, this assumption now imposes that the error terms are independent of the factors, the loadings and the covariates. One significant consequence of Assumption \hyperref[ER*]{{ER*}} is that $||\bs{\tilde{\varepsilon}}||_2 = \mathcal{O}_p(T^{\frac{3}{4}})$, which plays an important role in relaxing the requirement $T^2/n \rightarrow 0$ in Proposition \ref{AE}. 

The cross-sectional covariance matrix associated with the transformed error $\bs{\tilde{\varepsilon}}$ is $\bs{\Sigma}^{\frac{1}{2}}_{n} \bs{P}_{\bs{\mathscr{X}}} \bs{\Sigma}^{\frac{1}{2}}_{n}$. Therefore an additional assumption is required to manage the degree of dependence in this matrix. 
\begin{AssumptionAD}[Asymptotic Distribution]\label{AD} 
	There exists an $n \times n$ nonstochastic matrix $\bs{\Pi}$ with elements $\pi_{ij}$, which is uniformly bounded in absolute row and column sums, and such that, for any $\bs{\mathscr{X}}$, $|({\Sigma}^{\frac{1}{2}}_{n} {P}_{\bs{\mathscr{X}}} {\Sigma}^{\frac{1}{2}}_{n})_{ij}| \leq |\pi_{ij}|$ for all $i,j$. 
\end{AssumptionAD}
This assumption asserts that, irrespective of realisations of the covariates, the matrix $\bs{\Sigma}^{\frac{1}{2}}_{n} \bs{P}_{\bs{\mathscr{X}}} \bs{\Sigma}^{\frac{1}{2}}_{n}$ is dominated by a nonstochastic matrix which is uniformly bounded in absolute row and column sums.

\subsection{Asymptotic Distribution: $T/n \rightarrow c \geq 0$}
Where $T/n \rightarrow c > 0$, Assumption \hyperref[ER*]{ER*} alone will not be enough to apply Proposition \ref{AE} because in this case $||\bs{\hat{\theta}} - \bs{\theta}^0||_2 = \smallO_p(1)$ does not follow from Proposition \ref{CS}. In order to resolve this, the following additional assumption is imposed under which it is possible to obtain a faster rate of consistency.  
\begin{AssumptionBE}[Bounded Elements]\label{BE}
The sum of the absolute value of the off-diagonal elements of $\bs{\Sigma}_T$ are bounded by a constant, that is, $\sum^T_{t=1}\sum^T_{\tau\neq t}|({\Sigma}_T)_{t\tau}| < c_{\Sigma_T}$. 	
\end{AssumptionBE}
Assumption \hyperref[BE]{{BE}} places further limits on the degree of inter-temporal dependence. The precise origin of this assumption will become clear shortly. With this assumption in place, the following faster rate of consistency can be obtained. 
\begin{proposition}[Consistency -- Faster]\label{c1.5} 
	Under Assumptions \hyperref[AMD]{{MD}}, \hyperref[ACS]{{CS}},  \hyperref[ER*]{{ER*}} and \hyperref[BE]{{BE}},
	\begin{align}
		||\bs{\hat{\theta}} - \bs{\theta}^0||_2 = \mathcal{O}_p \left( \frac{T^{\frac{1}{4}}}{\sqrt{n}} \right). 
	\end{align}  
\end{proposition}

Proposition \ref{c1.5} refines the rate of consistency given in Proposition \ref{CS}, though does so under more stringent conditions. This result enables the derivation of Theorem \ref{Thrm1} presented below. 
\begin{theorem}[Asymptotic Distribution]\label{Thrm1} Under Assumptions
	\hyperref[AMD]{{MD}}, \hyperref[ACS]{{CS}}, \hyperref[AAE]{{AE}}, \hyperref[AD]{{AD}}, \hyperref[ER*]{{ER*}}, and \hyperref[BE]{{BE}}, as $T/n \rightarrow c$ with $c \in [0,K^{-1}]$,
\begin{align}
	\sqrt{nT}  (\bs{\hat{\theta}} - \bs{\theta}^0) + \bs{\Delta}^{-1} (\bs{\psi}^{(0)} + \bs{\psi}^{(1)}  + \bs{\psi}^{(2)}  + \bs{\psi}^{(3)}) 
	\xrightarrow{d}
	\mathcal{N}(\bs{0}, \bs{\Delta}^{-1} ( \bs{\Omega} + \bs{\Upsilon}^{(2)} + \bs{\Xi} + \bs{\bar{\Phi}} )\bs{\Delta}^{-1}),
	\label{first} 
\end{align} 
	where, \small
	\begin{align}
		\bs{\psi}^{(0)}
		\coloneqq  
		\frac{1}{\sqrt{nT}} 
		\begin{pmatrix}
			\textup{tr}( \bs{M}_{\bs{\tilde{\Lambda}^0}} \bs{\tilde{\Sigma}}_n ) \textup{tr}( \bs{G} \bs{\Sigma}_T ) \\
			\bs{0}_{K \times 1
		}	\end{pmatrix},
	\end{align}
	\begin{align}
	\bs{\psi}^{(1)}
	\coloneqq  
	\frac{1}{\sqrt{nT}} 
	\begin{pmatrix}
	\textup{tr}( \bs{\tilde{\Sigma}}_n ) (\textup{tr}( \bs{\Sigma}_T \bs{M}_{\bs{F}^0}  \bs{G} \bs{P}_{\bs{F}^0} )
		+
	\textup{tr}( \bs{P}_{\bs{F}^0} \bs{\Sigma}_T  \bs{G}  )) \\
	\bs{0}_{K \times 1
	}	\end{pmatrix},
	\end{align}
	\begin{align}
	{\psi}^{(2)}_{\kappa}
	\coloneqq 
	\frac{1}{\sqrt{nT}} \textup{tr}(\bs{\Sigma}_{T}) \textup{tr}( \bs{\tilde{\Sigma}}_n  \bs{M}_{\bs{\tilde{\Lambda}}^0}  \bs{{\tilde{Z}}}_{\kappa} \bs{F}^0 (\bs{F}^{0\top} \bs{F}^0)^{-1} (\bs{\tilde{\Lambda}}^{0\top} \bs{\tilde{\Lambda}}^0)^{-1} \bs{\tilde{\Lambda}}^{0\top} ),
	\end{align}
	\begin{align}
	{\psi}^{(3)}_{\kappa}
	\coloneqq 
	\frac{1}{\sqrt{nT}} \textup{tr}( \bs{\tilde{\Sigma}}_n) \textup{tr}( \bs{\Sigma}_{T} \bs{F}^0 (\bs{F}^{0\top} \bs{F}^0)^{-1} (\bs{\tilde{\Lambda}}^{0\top} \bs{\tilde{\Lambda}}^0)^{-1} \bs{\tilde{\Lambda}}^{0\top}  \bs{\tilde{Z}}_{\kappa}\bs{M}_{\bs{F}^0}),
\end{align}
\begin{align}
	\bs{\Omega}
	\coloneqq 
	\frac{1}{nT} \bs{\mathcal{\tilde{Z}}}^\top (\bs{M}_{\bs{F}^0} \otimes \bs{M}_{\bs{\tilde{\Lambda}}^0})   ( \bs{\Sigma}_T  \otimes  \bs{\tilde{\Sigma}}_n ) (\bs{M}_{\bs{F}^0} \otimes \bs{M}_{\bs{\tilde{\Lambda}}^0}) \bs{\mathcal{\tilde{Z}}}, \label{upup}
\end{align}
\begin{align}
	\bs{\Upsilon}^{(1)}
	\coloneqq 
	 \frac{1}{nT} \
	\begin{pmatrix}
		\textup{tr}(\bs{\tilde{\Sigma}}_n )\textup{tr}(\bs{G}\bs{\Sigma}_T\bs{G}^\top)  & \bs{0}_{1 \times K} \\
		\bs{0}_{K \times 1} 	& \bs{0}_{K \times K} \label{upupup}
	\end{pmatrix},
\end{align} 
\begin{align}
	\bs{\Upsilon}^{(2)}
	\coloneqq 
	\frac{1}{nT} \frac{1}{2}
	\begin{pmatrix}
		\textup{tr}(\bs{\tilde{\Sigma}}_n\bs{\tilde{\Sigma}}_n )(\textup{tr}(\bs{G}\bs{\Sigma}_T\bs{G}\bs{\Sigma}_T) + 2\textup{tr}(\bs{G}\bs{\Sigma}_T\bs{G}^\top\bs{\Sigma}_T) + \textup{tr}(\bs{G}^\top \bs{\Sigma}_T\bs{G}^\top \bs{\Sigma}_T)) & \bs{0}_{1 \times K} \\
		\bs{0}_{K \times 1} 	& \bs{0}_{K \times K}
	\end{pmatrix},
\end{align} 
\begin{align}
	\bs{\Xi}
	\coloneqq
	\frac{(\upsilon^{(4)} - 3)}{nT}	
	\begin{pmatrix}
		\textup{tr}(\textup{diagv}(\bs{\Sigma}^{\frac{1}{2}}_T \bs{G}\bs{\Sigma}^{\frac{1}{2}}_T)^\top \textup{diagv}(\bs{\Sigma}^{\frac{1}{2}}_T \bs{G}\bs{\Sigma}^{\frac{1}{2}}_T))\textup{tr}( \textup{diagv}(\bs{\Sigma}^{\frac{1}{2}}_n \bs{P}_{\bs{\mathscr{X}}} \bs{\Sigma}^{\frac{1}{2}}_n)^\top \textup{diagv}(\bs{\Sigma}^{\frac{1}{2}}_n \bs{P}_{\bs{\mathscr{X}}} \bs{\Sigma}^{\frac{1}{2}}_n))   & \bs{0}_{1 \times K} \\
		\bs{0}_{K \times 1} 	& \bs{0}_{K \times K}
	\end{pmatrix},
\end{align}
$\bs{\tilde{\Sigma}}_n \coloneqq \bs{Q}_{\bs{\mathscr{X}}}^\top \bs{\Sigma}_n \bs{Q}_{\bs{\mathscr{X}}}$, $\bs{\Delta} \coloneqq \bs{D} + \bs{\Upsilon}^{(1)}$, $\bs{\bar{\Phi}} \coloneqq \upsilon^{(3)} ( \bs{\Phi} + \bs{\Phi}^\top )/nT$, $\bs{\Phi} \coloneqq (\bs{\phi}, \bs{0}_{(K + 1) \times K})$, with $\bs{\phi}$ being a $(K+1) \times 1$ vector with $\kappa$-th element
\begin{align}
\phi_{\kappa} 
\coloneqq   \textup{vec}( (\textup{diagv}(  \bs{\Sigma}^{\frac{1}{2}}_T \bs{G} \bs{\Sigma}^{\frac{1}{2}}_T) \otimes \textup{diagv}( \bs{\Sigma}^{\frac{1}{2}}_n \bs{P}_{\bs{\mathscr{X}}} \bs{\Sigma}^{\frac{1}{2}}_n)   ))^\top \textup{vec}(\bs{\Sigma}_{n}^{\frac{1}{2}} \bs{Q} \bs{M}_{\bs{\tilde{\Lambda}^0}} \bs{\tilde{Z}}_{\kappa}\bs{M}_{\bs{F}^0}\bs{\Sigma}^{\frac{1}{2}}_{T}),
\end{align}
${\upsilon}^{(3)}$ and  ${\upsilon}^{(4)}$ being the third and fourth moments of $u_{it}$, respectively, and where $\bs{D}$ is defined in Proposition \ref{AE}. 
\end{theorem}
What is perhaps most significant in Theorem \ref{Thrm1} are the four bias terms $\bs{\psi}^{{(0)}}, \bs{\psi}^{{(1)}}, \bs{\psi}^{{(2)}}$ and $\bs{\psi}^{{(3)}}$. The first two of these, $\bs{\psi}^{{(0)}}$ and $\bs{\psi}^{{(1)}}$, arise due to the presence of the dynamic regressor and reflect the complex temporal interaction, represented by $\bs{G}$, that occurs with the covariance matrices $\bs{\Sigma}_T$ and $\bs{\tilde{\Sigma}}_n$, the factors and the loadings. Terms $\bs{\psi}^{{(2)}}$ and $\bs{\psi}^{{(3)}}$ arise due to cross-sectional and time series dependence and, as is discussed shortly, are notably absent in the case of identically and independently distributed errors. Studying  all these terms more closely shows the order of these biases to be:\footnote{Although random, strictly speaking the elements of the vectors $\bs{\psi}^{{(0)}}$ and $\bs{\psi}^{{(1)}}$ are bounded by constants and not just in probability.}
\begin{align}
	\bs{\psi}^{(0)} 
	&= \bs{\mathcal{O}}_p \left( \sqrt{\frac{T}{n}} \right), \label{b1}\\
	\bs{\psi}^{(1)} 
	&= \bs{\mathcal{O}}_p \left( \sqrt{\frac{T}{n}} \right), \\
	\bs{\psi}^{(2)} 
	&= \bs{\mathcal{O}}_p \left( \sqrt{\frac{T}{n}} \right), \\
	\bs{\psi}^{(3)} 
	&= \bs{\mathcal{O}}_p \left( \sqrt{\frac{T}{n}} \right),
\end{align}
which reveals something fundamental: projection of the entire model into the time dimension of the panel does not make the incidental problem in the cross-section disappear entirely, it instead shifts it into the time dimension, where it may interact with the extant problem in that dimension in complicated ways. 

Theorem \ref{Thrm1} also reveals the origin of the requirement $T^3/n \rightarrow 0$ in Proposition \ref{sqnTCS}, as well as the rationale behind Assumption \hyperref[BE]{{BE}}, both of which are a direct consequence of the order of the bias $\bs{\psi}^{{(0)}}$. For the first component of $\bs{\psi}^{{(0)}}$, it can be established that, under Assumptions \hyperref[AE]{{AE}} and \hyperref[ER*]{ER*},
\begin{align}
aT - b
\leq 
\textup{tr}( \bs{M}_{\bs{\tilde{\Lambda}^0}} \bs{\tilde{\Sigma}}_n ), \label{eqnb}
\end{align}
where $a$ and $b$ are positive constants.\footnote{See Appendix A.2.} The second component, $\text{tr}(\bs{G}\bs{\Sigma}_T)$, is a weighted summation of the lower triangular elements of $\bs{\Sigma}_T$, and, under Assumptions \hyperref[AMD]{{MD}} \hyperref[ER*]{{ER*}}, it can be established that 
\begin{align}
|\text{tr}(\bs{G}\bs{\Sigma}_T)|
\leq
T||\bs{G}||_2||\bs{\Sigma}_T||_2
= \mathcal{O}_p(T).
\end{align}
Indeed, without further restrictions, it is possible for this term to be exactly of that order, in which case 
\begin{align}
\bs{\psi}^{(0)} 
=
\mathcal{O}_p\left(\frac{T^{1.5}}{\sqrt{n}}\right). \label{bias}
\end{align}
This lays bare the origin of the requirement $T^3/n \rightarrow 0$ in Proposition \ref{sqnTCS}. Moreover, \eqref{bias} also suggests that, without further restrictions, rates of consistency faster than $\sqrt{T/n}$ should not be expected, even where the number of factors is known. The addition of Assumption \hyperref[BE]{{BE}} ensures $\text{tr}(\bs{G}\bs{\Sigma}_T) = \mathcal{O}_p\left( 1 \right)$, which reduces the order of $\bs{\psi}^{(0)}$ to that given in \eqref{b1}. The following subsections examine more closely $4$ special cases of Theorem \ref{Thrm1} that are of particular interest. 

\subsubsection{Diagonal $\bs{\Sigma}_T$}
A special case of Assumption \hyperref[BE]{BE} is where the off-diagonal elements of $\bs{\Sigma}_T$ are exactly zero, in which case the following corollary is obtained. 
\begin{corollary}[Diagonal $\bs{\Sigma}_T$]\label{corrr}
	Assume $\bs{\Sigma}_T$ is a positive diagonal matrix with bounded entries. Then, under the assumptions of Theorem \ref{Thrm1}, as $T/n \rightarrow c$ with $c \in [0,K^{-1}]$,
	\begin{align}
		\sqrt{nT}  (\bs{\hat{\theta}} - \bs{\theta}^0) + \bs{\Delta}^{-1} (\bs{\psi}^{(1)}  + \bs{\psi}^{(2)}  + \bs{\psi}^{(3)}) 
		\xrightarrow{d}
		\mathcal{N}(\bs{0}, \bs{\Delta}^{-1} ( \bs{\Omega} + \bs{\Upsilon}^{(2)} )\bs{\Delta}^{-1}),
		\label{seconda} 
	\end{align}
	where the nonzero element in $\bs{\Upsilon}^{(2)}$ simplifies to $2\textup{tr}(\bs{\tilde{\Sigma}}_n\bs{\tilde{\Sigma}}_n )\textup{tr}(\bs{G}\bs{\Sigma}_T\bs{\Sigma}_T\bs{G}^\top)$. 
\end{corollary}
When $\bs{\Sigma}_T$ is diagonal, the variance-covariance matrix is greatly simplified and the bias term $\bs{\psi}^{(0)}$ is identically zero.\footnote{This can be seen by noticing that $\text{tr}(\bs{G} \bs{\Sigma}_{T}) 	= \sum^{T-1}_{t = 1} \sum^{t}_{\tau = 1} \alpha^{\tau - 1}	(\bs{\Sigma}_{T})_{(t+1,t+1-\tau)}$, which is a summation over the lower triangular elements of $\bs{\Sigma}_{T}$. Therefore, as long as $\bs{\Sigma}_T$ is upper triangular, or indeed diagonal, this will be exactly zero.} As a consequence, the estimator is $\sqrt{nT}$ consistent as $T/n \rightarrow c$ with $c \in [0,K^{-1}]$, and is in fact always at least $\sqrt{n}$ consistent irrespective of $T$. Moreover, where $T/n \rightarrow 0$, the estimator is asymptotically unbiased, though, in the event that $T/n \rightarrow c > 0$, a bias of order $\sqrt{T/n}$ is present. 

\subsubsection{IID Errors}
Another corollary of Theorem \ref{Thrm1} is obtained when the errors are identically and independently distributed. In this case $\bs{\psi}^{(2)} = \bs{\psi}^{(3)} = \bs{0}_{(K+1) \times 1}$ because where $\bs{\Sigma}_T \propto \bs{I}_T$, the orthogonal projector $\bs{M}_{\bs{F}^0}$ directly multiplies with $\bs{F}^{0}$ and, similarly, when $\bs{\Sigma}_n \propto \bs{I}_n$, $\bs{M}_{\bs{\tilde{\Lambda}^0}}$ directly multiples with $\bs{\tilde{\Lambda}}^0$. 
\begin{corollary}[IID]\label{homsa} 
	Assume $\bs{\Sigma}_n = \sigma_0^2 \bs{I}_n$ and $\bs{\Sigma}_T = \sigma_0^2 \bs{I}_T$. Then, under the assumptions of Theorem \ref{Thrm1}, as $T/n \rightarrow c$ with $c \in [0,K^{-1}]$,
	\begin{align}
		\sqrt{nT}  (\bs{\hat{\theta}} - \bs{\theta}^0) + \bs{\Delta}^{-1} \bs{\psi}^{(1)} 
		\xrightarrow{d}
		\mathcal{N}(\bs{0}, \bs{\Delta}^{-1} ( \bs{\Omega} + \bs{\Upsilon}^{(2)} )\bs{\Delta}^{-1}),
		\label{homb} 
	\end{align}
	where the nonzero element in $\bs{\Upsilon}^{(2)}$ simplifies to $2TK\sigma_0^4\textup{tr}(\bs{G}\bs{G}^\top)$.
\end{corollary}
In this exceptional case the only remaining bias $\bs{\psi}^{(1)}$ reduces to 
\begin{align}
	{\psi}_{\sigma}^{(1)}
	&\coloneqq 
	\frac{\sigma^2_0}{\sqrt{nT}}
	\textup{tr}( \bs{P}_{\bs{\mathscr{X}}} ) \textup{tr}( \bs{G} \bs{P}_{\bs{F}^0} ).
\end{align}  
This, in fact, is a generalisation of the bias described in \citep{nickell}. To see this, recall that the model of individual effects nests as a special case of interactive fixed effects in which a single heterogeneous loading vector is interacted with a constant factor:
\begin{align}
	\bs{\lambda}_{IFE}^0 
	\coloneqq 
	\begin{pmatrix}
		\lambda_1^0 \\
		\vdots \\
		\lambda_{n}^0
	\end{pmatrix},\ 
	\bs{F}_{IFE}^0
	\coloneqq
	\bs{\iota}_{T},
\end{align}
where $\bs{\iota}_{T}$ is a $T \times 1$ vector of ones. In this case $\bs{P}_{\bs{F}_{IFE}^0} = \frac{1}{T}\bs{\iota}_{T}\bs{\iota}_T^\top$, and therefore,
\begin{align}
	{\psi}_{\sigma}^{(2)}
	&= 	
	\frac{\sigma^2_0}{\sqrt{nT}} \frac{1}{T}
	\textup{tr}( \bs{P}_{\bs{\mathscr{X}}})\textup{tr}( \bs{G} \bs{\iota}_{T}\bs{\iota}_T^\top).
\end{align} 
Notice the structure of $\bs{G}$:
\begin{align}
	\bs{G} \bs{\iota}_{T}\bs{\iota}_T^\top
	=
	\begin{pmatrix}
		\vdots                & \vdots                & \vdots  \\
		1 + \alpha^0 + (\alpha^0)^2 & 1 + \alpha^0 + (\alpha^0)^2 & 1 + \alpha^0 + (\alpha^0)^2 \\
		1 + \alpha^0            & 1 + \alpha^0            & 1 + \alpha^0 \\
		1                     & 1                     & 1 \\
		0                     & 0                     & 0
	\end{pmatrix},
\end{align}
and so $\textup{tr}( \bs{G} \bs{\iota}_{T}\bs{\iota}_T^\top) = \sum^{T-1}_{t=1}\sum^t_{\tau = 1} (\alpha^0)^{\tau-1}$. A bit of algebra reveals that\footnote{See Appendix A.3.} 
\begin{align}
	\sum^{T-1}_{t=1}\sum^t_{\tau = 1} (\alpha^0)^{\tau-1}
	= 
	\frac{T}{(1-\alpha^0)} \left( 1 - \frac{1}{T} \frac{(1-(\alpha^0)^T)}{1-\alpha^0} \right). \label{nikf}
\end{align}
Now, since the trace of a projector is equal to its rank, $\text{tr}(\bs{P}_{\bs{\mathscr{X}}}) = TK$, and the following expression is obtained:
\begin{align}
	\psi^{(1)}_{\sigma}
	=
	\sqrt{\frac{T}{n}} \frac{K}{(1-\alpha)} \left( 1 - \frac{1}{T} \frac{(1-\alpha^T)}{1-\alpha} \right). \label{thisd}
\end{align}
This reveals a simple expression for the bias. Notice the significance of the transformation $\bs{Q}_{\bs{\mathscr{X}}}$: it reduces the rank of the cross-sectional covariance matrix to $TK$. Without the transformation $\text{tr}(\bs{\tilde{\Sigma}}_n) = \text{tr}(\bs{{\Sigma}}_n) = \text{tr}(\bs{I}_n) = n$, and so 
\begin{align}
	\psi^{(1)}_{\sigma}
	=
	\sqrt{\frac{n}{T}} \frac{1}{(1-\alpha)} \left( 1 - \frac{1}{T} \frac{(1-\alpha^T)}{1-\alpha} \right), \label{niiiff}
\end{align}
which matches (up to scale by $\sqrt{nT}$) exactly expression (27) derived in \cite{nickell}. This again highlights the fact that transforming the model by $\bs{Q}_{\bs{\mathscr{X}}}$ does not eliminate all traces of the incidental parameter problem that would have existed in the cross-section. It simply transfers it to the time dimension where, as exemplified by comparing \eqref{thisd} and \eqref{niiiff},  it will likely manifest itself in similar ways. 

\subsubsection{Standard Normality}\label{snornm}
As would be expected, under standard normality of the errors particularly simple expressions are obtained. 
\begin{AssumptionSN}[Standard Normality]\label{ASN} Conditional on the covariates, the factors and the loadings, the elements of the $n \times T$ matrix $\bs{\varepsilon}$ are drawn from independent standard normal distributions.
\end{AssumptionSN}
\begin{corollary}[Standard Normality]\label{Thrm3} 
	Under Assumptions \hyperref[AMD]{{MD}}, \hyperref[ACS]{{CS}},  \hyperref[AAE]{{AE}}, \hyperref[AD]{{AD}}, and \hyperref[ASN]{{SN}}, as $T/n \rightarrow c$ with $c \in [0,K^{-1}]$,
	\begin{align}
		\sqrt{nT}  (\bs{\hat{\theta}} - \bs{\theta}^0) - \bs{\Delta}^{-1} \bs{\psi}_{SN}
		\xrightarrow{d}
		\mathcal{N}(\bs{0}, \bs{\Delta}^{-1}),
		\label{first} 
	\end{align} 
	where \small
	\begin{align}
		\bs{\psi}_{SN}
		&\coloneqq 
		K\sqrt{\frac{T}{n}} 
		\begin{pmatrix}
			\textup{tr}( \bs{G}\bs{P}_{\bs{F}^0})  \\
			\bs{0}_{K \times 1
		}	\end{pmatrix},
	\end{align}
	and $\bs{\Delta}$ is defined in Theorem \ref{Thrm1}. 
\end{corollary}
Following the same steps used to obtain \eqref{thisd}, a simple expression for $\bs{\psi}_{SN}$ can be derived under individual effects. In general, however, using the inequality $\text{tr}(\bs{A}) \leq \text{rank}(\bs{A})||\bs{A}||_1$, and noticing that $||\bs{G}||_1 = \sum^{T-1}_{t=1} (\alpha^0)^{t-1}$, one obtains the following bound:
\begin{align}
	\sqrt{\frac{T}{n}} K |\textup{tr}( \bs{G}\bs{P}_{\bs{F}^0})|
	\leq 
	\sqrt{\frac{T}{n}} K R^0\frac{ (1 - (\alpha^0)^T) }{1 - \alpha^0},
\end{align}
and hence the bias remains comparable, even where the structure of $\bs{P}_{\bs{F}^0}$ is left unrestricted. 

\subsubsection{Static Model}
\begin{corollary}[Static Model]\label{stats} 
	In the absence of a dynamic regressor, then, under the assumptions of Theorem \ref{Thrm1}, as $T/n \rightarrow c$ with $c \in [0,K^{-1}]$,
	\begin{align}
		\sqrt{nT}  (\bs{\hat{\beta}} - \bs{\beta}^0) + \bs{D}^{-1} (\bs{\psi}^{(2)}  + \bs{\psi}^{(3)}) 
		\xrightarrow{d}
		\mathcal{N}(\bs{0}, \bs{D}^{-1} \bs{\Omega} \bs{D}^{-1}).
		\label{thirdaa} 
	\end{align} 
\end{corollary}
This result is the analogue of Theorem 3 in \cite{bai_panel_2009} with the important difference being the order of the bias terms, which in this case are both of order $\sqrt{T/n}$. Moreover, similarly to Corollary \ref{corrr}, this result implies that, in the absence of dynamics, the estimator must be at least $\sqrt{n}$ consistent irrespective of $T$. 

\subsection{Asymptotic Distribution: Fixed $T$}
This section presents the main result of this paper in the form of the following theorem.
\begin{theorem}[Fixed $T$]\label{fixedT}
Under Assumptions \hyperref[AMD]{{MD}}, \hyperref[ACS]{{CS}}, \hyperref[AAE]{{AE}}, \hyperref[AD]{{AD}}, and \hyperref[ER*]{{ER*}}, with $T$ fixed and $n \rightarrow \infty$,
	\begin{align}
		\sqrt{nT}  (\bs{\hat{\theta}} - \bs{\theta}^0) 
		\xrightarrow{d}
		\mathcal{N}(\bs{0}, \bs{D}^{-1} \bs{\Omega} \bs{D}^{-1}).
		\label{third} 
	\end{align}
\end{theorem}
Theorem \ref{fixedT} demonstrates that with $T$ fixed, the estimator is asymptotically unbiased in the presence of cross-sectional dependence, serial dependence, and with the inclusion of dynamic regressors. Notice also that Assumption \hyperref[BE]{{BE}} is not required to obtain this result because the bias $\bs{\psi}^{(0)} = \smallO_p(1)$ when $T$ is fixed.


\begin{remark}
Although the result for the static model is presented as a corollary of Theorem \ref{Thrm1}, where there is are no lagged outcome $\bs{\psi}^{(0)}$ does not appear, and so this result can be obtained without requiring Assumption \hyperref[BE]{{BE}}. 
\end{remark}

\begin{remark}
While the biases $\bs{\psi}^{(1)}, \bs{\psi}^{(2)}$ and $\bs{\psi}^{(3)}$ have analogues described in \cite{bai_panel_2009}, \cite{moon_linear_2015}, and  \cite{moon_dynamic_nodate}, $\bs{\psi}^{(0)}$ does not. That is because the former two of these papers study the static model, while the latter derives the asymptotic distribution under the assumption that the errors are independent across time, whereby $\bs{\Sigma}_T$ is diagonal, and $\bs{\psi}^{(0)}$ does not appear. 
\end{remark}

\section{Further Matter}\label{furmat}

\subsection{Faster Rates of Consistency with ${R} \geq R^0$.}\label{stdn}
Some of the situations discussed in Section \ref{asydist} also give rise to faster rates of consistency with $R \geq R^0$. This section serves to highlight a few of these cases. 
\begin{proposition}[Consistency -- Static Model]\label{c2}
	In the absence of a dynamic regressor, under Assumptions \hyperref[AMD]{{MD}}, \hyperref[ACS]{{CS}}, and \hyperref[ER*]{{ER*}},
	\begin{align}
		||\bs{\hat{\beta}} - \bs{\beta}^0||_2 = \mathcal{O}_p \left( \frac{T^{\frac{1}{4}}}{\sqrt{n}} \right). 
	\end{align} 
\end{proposition}
At the core of this result is the fact that under Assumption \hyperref[ER*]{{ER*}} $||\bs{\tilde{\varepsilon}}||_2 = \mathcal{O}_{p}(T^{\frac{3}{4}})$. Because, trivially, it is also the case that $||\bs{\tilde{\varepsilon}}||_2 \leq ||\bs{{\varepsilon}}||_2$, the estimator must also at least be consistent at the rate $\frac{1}{\sqrt{T}}$, which is straightforward to obtain following Theorem 4.1 in \cite{moon_linear_2015}. 

Owing to the fact that the standard normal distribution is invariant to orthogonal transformations, especially favourable rates of consistency can be achieved in this case. Key to showing this is the following result.
\begin{proposition}\label{normalnorm}
	Under Assumption \hyperref[ASN]{{SN}}, $||\bs{\tilde{\varepsilon}}||_2 = \mathcal{O}_p(\sqrt{T})$. 
\end{proposition}
\begin{pfff}\
	Since the normal distribution is invariant to orthogonal transformation, it follows that $\bs{Q}_{\bs{\mathscr{X}}}^\top \bs{\varepsilon}$ is a $TK \times T$ matrix with independent standard normal entries. \cite{latala} shows that such a matrix will be $\mathcal{O}_p(\max\{\sqrt{TK},\sqrt{T}\}) = \mathcal{O}_p(\sqrt{T})$. 
\end{pfff}\\
Using Proposition \ref{normalnorm}, a faster rate of consistency can thus be obtained. 
\begin{proposition}[Consistency -- Standard Normality]\label{c3} 
	Under Assumptions \hyperref[AMD]{{MD}}, \hyperref[ACS]{{CS}}, and \hyperref[ASN]{{SN}},
	\begin{align}
		||\bs{\hat{\theta}} - \bs{\theta}^0||_2 = \mathcal{O}_p \left( \frac{1}{\sqrt{n}} \right). 
	\end{align} 
\end{proposition}
This result demonstrates that, under standard normality, and with $R \geq R^0$, the rate of consistency is in fact independent of $T$.

\subsection{Low Rank Covariates}
Low rank covariates often appear in applied work, with obvious examples being those that are either time or cross-sectionally invariant. In models with interactive effects, identifying the coefficients associated with these covariates can be challenging since they present another low rank structure in the model, in addition to the factor term. Mirroring the result obtained in \cite{moon_dynamic_nodate}, where such covariates are present it is, however, still possible to obtain consistent estimates under appropriate conditions. Let $\bs{\vartheta}$ denote a reordering of the parameter vector $\bs{\theta}$ such that the first $K_{\tx{L}}$ coefficients, indexed $l = 1,..., K_{\tx{L}}$, are those associated with low rank regressors, and the remaining $K_{\tx{H}}$ coefficients, indexed $h = 1,...,K_{\tx{H}}$, denote those associated with the regressors which have full rank. For simplicity it is assumed that the low rank regressors have rank $1$, though the following results extend naturally to the more general case. The $l$-th low rank covariate can be decomposed as $\bs{X}_{l} = \bs{v}_{l}\bs{w}^\top_{l}$, with $\bs{v}_l$ and $\bs{w}_l$ being $n \times 1$ and $T \times 1$ vectors, respectively. These vectors can then be gathered into the matrices $\bs{\mathcal{V}} \coloneqq (\bs{v}_1,...,\bs{v}_{K_\tx{L}})$ and $\bs{\mathcal{W}} \coloneqq (\bs{w}_1,...,\bs{w}_{K_{\tx{L}}})$. When some of the covariates are low rank, special care must be taken in the construction of $\bs{\mathscr{X}}$. In this case $\bs{\mathscr{X}}$ can be constructed as $ (\bs{\mathcal{V}}, \bs{X}_1,...,\bs{X}_{K_{\tx{H}}} ) $ to ensure that $\bs{\mathscr{X}}^\top \bs{\mathscr{X}}$ is invertible. Let $\bs{\tilde{\mathcal{V}}} = \bs{Q}_{\bs{\mathscr{X}}}^\top \bs{\mathcal{V}}$ and $\bs{\delta}_{\tx{H}} \cdot \bs{\tilde{{Z}}}_{\tx{H}} \coloneqq \sum^{K_{\tx{H}}}_{\kappa = 1} \delta_{\kappa} \bs{\tilde{Z}}_{\kappa}$.
\begin{AssumptionLR}[Low Rank]\label{ALR}\color{white}.\color{black}\
	\begin{enumerate}[label=\textup{(\roman*)}]
		\item $\min_{\bs{\delta}_{\tx{H}} \in \mathbb{R}^{K_{\tx{H}}}: ||\bs{\delta}_{\tx{H}}||_2 = 1} \sum^T_{r = R + R^0 + K_{\tx{L}} + 1} \mu_r \left( \frac{1}{nT} (\bs{\delta}_{\tx{H}} \cdot \bs{\tilde{{Z}}}_{\tx{H}})^\top (\bs{\delta}_{\tx{H}} \cdot \bs{\tilde{{Z}}}_{\tx{H}}) \right) \geq b > 0$. \label{ALRi}	
		\item There exists a constant $c > 0$ such that $\frac{1}{n}\bs{\tilde{\Lambda}}^{0\top}\bs{M}_{\bs{\tilde{\mathcal{V}}}}\bs{\tilde{\Lambda}}^0 > c \bs{I}_{R^0}$ and $\frac{1}{T}\bs{F}^{0\top}\bs{M}_{\bs{\mathcal{W}}}\bs{F}^0 > c \bs{I}_{R^0}$, w.p.a.1. \label{ALRii}		
	\end{enumerate} 
\end{AssumptionLR}
Assumption \hyperref[ACS]{{LR}} is analogous to Assumption 4(ii) in \cite{moon_dynamic_nodate} and requires what amounts to a strengthening of Assumption \hyperref[ACSii]{{CS(ii)}}, and an additional condition to ensure that the low rank regressors are sufficiently distinct from the factors and the transformed loadings so as to be able to distinguish one from the other. Here however, special care must be taken with Assumption \hyperref[ACS]{{LR(ii)}} because 
\begin{align}
\frac{1}{n}\bs{\tilde{\Lambda}}^{0\top}\bs{M}_{\bs{\tilde{\mathcal{V}}}}\bs{\tilde{\Lambda}}^0
=
\frac{1}{n}\bs{{\Lambda}}^{0\top}( \bs{P}_{\bs{{\mathscr{X}}}} - \bs{P}_{\bs{{\mathcal{V}}}}) \bs{{\Lambda}}^0.
\end{align}
Since the transforming the model by $\bs{Q}_{\bs{\mathscr{X}}}^\top$ has the effect of projecting the model into the column space of the covariates, it is not enough that the loadings be distinct from each $\bs{v}_{l}$, as in \cite{moon_dynamic_nodate}. In this context what is required is that the projection of the loadings onto the column space of the all the covariates is different from the projection onto the column space of just the low rank covariates, which, clearly, will require there to be some high rank model covariates. 
\begin{proposition}[Consistency -- Low Rank]\label{c4} 
	Under Assumptions \hyperref[AMD]{{MD}}, \hyperref[AAE]{{AE}}, \hyperref[ADE]{{ER}},  and \hyperref[ACS]{{LR}}, \footnote{Faster rates will also be possible in the circumstances discussed in Section \ref{stdn}.}
	\begin{align}
	||\bs{\hat{\vartheta}} - \bs{\vartheta}^0||_2 = \mathcal{O}_p \left( \sqrt{\frac{T}{n}} \right).
	\end{align} 
\end{proposition}

\subsection{Estimating the Number of Factors}\label{estfac}
Results established in Sections \ref{consub} and \ref{stdn} demonstrate that in many instances the estimator will remain consistent with the number of factors overestimated. However, since overestimation of the number of factors will typically lead to a loss of efficiency in finite samples, it is desirable to input the correct number of factors. One approach to detecting this number involves first estimating the coefficients with the number of factors overestimated, and using these estimates to construct a pure factor model. Then, methods devised to detect the number of factors in a pure factor model can be applied. Examples of these detection methods include \cite{baifac}, \cite{onatski} and \cite{horenstein}. This section focuses on one of these, the eigenvalue ratio test of \cite{horenstein}, and considers how, after transforming the model, this method can be applied to detect the number of factors with $T$ fixed.

More generally, however, this section seeks to make two points. First, after having transformed the model, other results which exist in the literature for the large $n$, large $T$ setting may be ported to that with $T$ fixed, potentially with the additional benefit of relaxing assumptions regarding dependence in the errors. Second, in situations where factors exist in the error term which are uncorrelated with the covariates, alongside those which are correlated, transforming the model and detecting the number of factors may lead to efficiency gains, since only the number of factors which are correlated with the error term need be inputted into the estimation procedure. Let 
\begin{align}
	\mu^*_{r}  
	\coloneqq \mu_{r} \left( \frac{1}{nT}\left(\bs{\tilde{Y}} \bs{S}(\hat{\alpha})- \sum^{K}_{\kappa=1}\hat{\beta}_\kappa\bs{\tilde{X}}_\kappa\right)^\top \left(\bs{\tilde{Y}} \bs{S}(\hat{\alpha})- \sum^{K}_{\kappa=1}\hat{\beta}_\kappa\bs{\tilde{X}}_\kappa\right) + \frac{1}{n} \bs{I}_T \right), \label{err}
\end{align}
that is, $\mu^*_{r}$ is the $r$-th largest eigenvalue of the right-hand side matrix. Then define
\begin{align}
	\textup{EigR}(r)
	\coloneqq
	\frac{\mu^*_{r}}{\mu^*_{r+1}}\ \text{for}\ r = 1,...,T-1.
\end{align} 
The main modification here from \cite{horenstein}'s original specification is the addition of the matrix $\frac{1}{n}\bs{I}_T$. This is added because, unlike in the original setting, where covariates are present the eigenvalues in \eqref{err} need not be strictly positive, and there is a non-zero probability that some of them are exactly zero. The addition of the identity matrix ensures that this cannot happen and, because the eigenvalues are demonstrated to converge at a rate of $1/n$ or slower, this does not impact the properties of the test. 

\begin{proposition} 
	Under Assumptions \hyperref[AMD]{{MD}},  \hyperref[ACS]{{CS}} and \hyperref[ER*]{{ER*}}, as $T/n \rightarrow 0$, \label{prop7}
	\begin{align}
		\textup{Pr}\left(\max_{1 \leq r \leq T} \mu^*_{r} = R^0\right) \rightarrow 1.
	\end{align}
\end{proposition}

\subsection{Balestra and Nerlove's Approach}\label{bn}
As mentioned previously, though often still yielding consistent estimates, 
using the PC estimator with the number of factors inputted $R$ exceeding the true number of factors $R^0$ will result in a loss of efficiency in finite samples. The estimator $\bs{\hat{\theta}}$ studied thus far in this paper treats the initial condition $\bs{y}_0 \bs{s}^\top(\alpha)$ as an additional parameter and, as a consequence, results in another factor appearing in the error term. An alternative approach which does not generate this additional factor is to follow \cite{balestra_nerlove} and include the projection of the lagged outcome onto the column space of the exogenous variables as an additional explanatory variable on the right hand-side of the outcome equation. This approach is naturally wedded to this paper's, since projecting lagged outcomes embeds them in the $TK$-dimensional space spanned by the columns of $\bs{\mathscr{X}}$. Consider the following outcome equation:
\begin{align}
	\bs{{Y}}^c
	=&\ 
	\alpha \bs{{Y}}_L^c
	+ 
	\sum^K_{k=1} \beta_k {\bs{X}}_k^c
	+
	\bs{{\Lambda}}^* \bs{{F}}^{*c\top}
	+
	\bs{{\varepsilon}}^c, \label{bnest}
\end{align}
where $\bs{Y}^c_L \coloneqq (\bs{y}_{1},...,\bs{y}_{T-1})$, and the matrices $\bs{Y}^c, \bs{X}_k^c, \bs{{\Lambda}}^* \bs{{F}}^{*c\top}$ and $\bs{\varepsilon}^c$ are $n \times T^c$, with $T^c \coloneqq T-1$. Clearly the trade off in adopting this approach is that, while no longer generating an additional factor, this does lead to the loss of a time period of data. Define $\bs{\mathscr{X}}^c \coloneqq (\bs{X}_1^c,...,\bs{X}_K^c)$ and $\bs{Q}_{\bs{\mathscr{X}}^c} \coloneqq  \bs{\mathscr{X}}^c (\bs{\mathscr{X}}^{c^\top} \bs{\mathscr{X}}^c)^{-1}$. Then, using $\sim$ to indicate transformed variables, as previously, consider the alternate objective function
\begin{align}
	{Q}^c(\bs{\theta})
	&\coloneqq 
	\frac{1}{nT^c}\text{tr}\left(\left(\bs{\tilde{Y}}^c - \sum^{K+1}_{\kappa =1} \theta_\kappa \bs{\tilde{Z}}_\kappa^c  - \bs{\tilde{\Lambda}}^* \bs{F}^{*c\top} \right)^\top \left(\bs{\tilde{Y}}^c - \sum^{K+1}_{\kappa =1} \theta_\kappa \bs{\tilde{Z}}_\kappa^c  - \bs{\tilde{\Lambda}}^* \bs{F}^{*c\top} \right)\right), 
\end{align}
	where $\bs{\tilde{Z}}_1^c \coloneqq \bs{Q}_{\bs{\mathscr{X}}^c}^\top \bs{{Y}}_L^c$ and $\bs{\tilde{Z}}_\kappa^c \coloneqq \bs{\tilde{X}}_\kappa^c$ for $\kappa = 2 ,..., K+1$. An alternative estimator $\bs{\hat{\theta}}_{BN}$ may then be defined as
\begin{align}
	\bs{\hat{\theta}}_{BN}
	\coloneqq
	\argmin_{\bs{\theta} \in {\Theta}} 
	{Q}^c(\bs{\theta}).
\end{align}
This estimator retains all of the essential properties of the $\bs{\hat{\theta}}$, including fixed $T$ consistency and an analogous asymptotic distribution. Moreover, this approach is especially appealing since it involves simply transforming the data by $\bs{Q}_{\bs{\mathscr{X}}}$ and then applying the usual PC estimator with no other modifications. 

\begin{remark}
	The estimation approach proposed in this paper shares a close kinship with the procedure suggested by \cite{chamberlain} for short panels with individual effects. In the present context, this could be understood as decomposing $\bs{\Lambda} = \bs{P}_{\bs{\mathscr{X}}} \bs{\Lambda}  + \bs{M}_{\bs{\mathscr{X}}} \bs{\Lambda} \eqqcolon \bs{\mathscr{X}} \bs{\Gamma}  + \bs{e}$, where $\bs{\Gamma}$ is a $TK \times T$ parameter to be estimated, and $\bs{e}$ is subsumed into the error term. \cite{chamberlain} suggests a minimum distance approach to jointly estimate $\bs{\theta}$ and $\bs{\Gamma}$, however, if one instead applies least squares, then concentrating out $\bs{\Gamma}$ and minimising with respect to the factors will yield an identical estimator.
\end{remark}

\section{Monte Carlo Simulations}\label{sims}
This section provides simulation results which highlight the different properties of the PC estimator when applied to the original and transformed models. In the following design the factors and loadings are both generated independently from standard normal distributions and the true number of factors is set equal to $2$; i.e. $R^0=2$. Two covariates are generated: $\bs{X}_1 = \bs{\Lambda} \bs{F}^\top + \bs{\eta}$, where $\bs{\eta}$ has elements drawn independently from a standard normal distribution, and $\bs{X}_2$, 
which is also drawn from a standard normal. The entries of the error $\bs{\varepsilon}$ are generated as $\bs{\Sigma}^{\frac{1}{2}}_n \bs{U} \bs{\Sigma}^{\frac{1}{2}}_T$, where the elements of $\bs{U}$ are independently drawn from a standard normal distribution, and $\bs{\Sigma}_n$ and $\bs{\Sigma}_T$ are diagonal matrices with elements drawn uniformly between $0.5$ and $2.5$. The number of Monte Carlo replications is 10000. Tables \ref{T1a} -- \ref{T1c} display the bias and the standard error of the standard least squares estimator (LS), the principal component estimator applied to the original model (PC), the approach described in Section \ref{bn} (BN) and the PC estimator applied to the transformed model (QPC).
\begin{table}[H]
	\renewcommand{\thetable}{\arabic{table}a}
	\caption{Bias (SE) $\alpha$}\label{T1a}
	\begin{adjustbox}{width=15cm,center} 
		\begin{tabular}{|c|rrr|rrr|rrr|rrr|}\cline{1-13} 
			& \multicolumn{3}{c|}{LS} & \multicolumn{3}{c|}{PC}   & \multicolumn{3}{c|}{BN} & \multicolumn{3}{c|}{QPC} \\ \cline{1-13} 
			\multicolumn{1}{|c|}{$n\ \backslash \ T$}   & \multicolumn{1}{c}{$6$} & \multicolumn{1}{c}{$9$} & \multicolumn{1}{c|}{$12$} &  \multicolumn{1}{c}{$6$} & \multicolumn{1}{c}{$9$} & \multicolumn{1}{c|}{$12$} &  \multicolumn{1}{c}{$6$} & \multicolumn{1}{c}{$9$} & \multicolumn{1}{c|}{$12$} &  \multicolumn{1}{c}{$6$} & \multicolumn{1}{c}{$9$} & \multicolumn{1}{c|}{$12$} \\ \cline{1-13} 
			30   & \bf{-0.003} & \bf{-0.002} &\bf{-0.001} & \bf{-0.106} &\bf{-0.005} & \bf{-0.002} & \bf{-0.043} & \bf{-0.004} & \bf{-0.003} & \bf{-0.105} & \bf{-0.007} & \bf{-0.002} \\
			     & (0.060)     & (0.040)     & (0.026)    & (0.222)     & (0.054)    & (0.040)     & (0.173)     & (0.054)     & (0.040)     & (0.276)     & (0.066)     & (0.041)     \\
			60   & \bf{-0.002} & \bf{-0.001} &\bf{-0.001} & \bf{-0.086} &\bf{-0.007} & \bf{-0.001} & \bf{-0.013} & \bf{-0.003} & \bf{-0.001} & \bf{-0.023} & \bf{-0.005} & \bf{-0.001} \\
		         & (0.038)     & (0.031)     & (0.026)    & (0.184)     & (0.049)    & (0.030)     & (0.093)     & (0.043)     & (0.030)     & (0.120)     & (0.050)     & (0.029)     \\
			90   & \bf{-0.001} & \bf{-0.001} & \bf{0.000} & \bf{-0.191} &\bf{-0.004} & \bf{-0.001} & \bf{-0.010} & \bf{-0.001} & \bf{-0.001} & \bf{-0.014} & \bf{-0.002} & \bf{-0.001} \\
			     & (0.031)     & (0.025)     & (0.021)    & (0.252)     & (0.041)    & (0.026)     & (0.077)     & (0.035)     & (0.026)     & (0.086)     & (0.039)     & (0.025)     \\
			150  & \bf{-0.001} & \bf{0.000}  & \bf{0.000} & \bf{-0.237} &\bf{-0.003} & \bf{-0.001} & \bf{-0.005} & \bf{-0.001} & \bf{0.000}  & \bf{-0.008} & \bf{-0.001} & \bf{0.000}  \\
			     & (0.027)     & (0.019)     & (0.017)    & (0.260)     & (0.034)    & (0.023)     & (0.067)     & (0.026)     & (0.022)     & (0.071)     & (0.028)     & (0.020)     \\
			300  & \bf{0.000}  & \bf{0.000}  & \bf{0.000} & \bf{-0.085} &\bf{-0.003} & \bf{-0.001} & \bf{-0.002} & \bf{-0.001} & \bf{0.000}  & \bf{-0.002} & \bf{-0.001} & \bf{0.000}  \\ 
			     & (0.017)     & (0.014)     & (0.012)    & (0.175)     & (0.030)    & (0.020)     & (0.034)     & (0.020)     & (0.015)     & (0.034)     & (0.051)     & (0.015)     \\ \cline{1-13} 
		\end{tabular}
	\end{adjustbox}
\end{table}

\begin{table}[H]
	\addtocounter{table}{-1}
	\renewcommand{\thetable}{\arabic{table}b}
	\caption{Bias (SE) $\beta_1$}\label{T1b}
	\begin{adjustbox}{width=15cm,center} 
		\begin{tabular}{|c|rrr|rrr|rrr|rrr|}\cline{1-13} 
			& \multicolumn{3}{c|}{LS} & \multicolumn{3}{c|}{PC}   & \multicolumn{3}{c|}{BN} & \multicolumn{3}{c|}{QPC} \\ \cline{1-13} 
			\multicolumn{1}{|c|}{$n\ \backslash \ T$}   & \multicolumn{1}{c}{$6$} & \multicolumn{1}{c}{$9$} & \multicolumn{1}{c|}{$12$} &  \multicolumn{1}{c}{$6$} & \multicolumn{1}{c}{$9$} & \multicolumn{1}{c|}{$12$} &  \multicolumn{1}{c}{$6$} & \multicolumn{1}{c}{$9$} & \multicolumn{1}{c|}{$12$} &  \multicolumn{1}{c}{$6$} & \multicolumn{1}{c}{$9$} & \multicolumn{1}{c|}{$12$} \\ \cline{1-13} 
			30   & \bf{0.473}  & \bf{0.480}  & \bf{0.486} & \bf{0.218}  & \bf{0.064} & \bf{0.042}  & \bf{0.154}  & \bf{0.041}  & \bf{0.031}  & \bf{0.126}  & \bf{0.057}  & \bf{0.047} \\
			     & (0.138)     & (0.110)     & (0.097)    & (0.242)     & (0.106)    & (0.082)     & (0.240)     & (0.097)     & (0.076)     & (0.262)     & (0.093)     & (0.073)     \\
			60   & \bf{0.475}  & \bf{0.483}  & \bf{0.486} & \bf{0.129}  & \bf{0.039} & \bf{0.027}  & \bf{0.054}  & \bf{0.014}  & \bf{0.009}  & \bf{0.040}  & \bf{0.012}  & \bf{0.009}  \\
			     & (0.123)     & (0.101)     & (0.087)    & (0.189)     & (0.077)    & (0.055)     & (0.141)     & (0.063)     & (0.046)     & (0.135)     & (0.063)     & (0.047)     \\
			90   & \bf{0.476}  & \bf{0.484}  & \bf{0.487} & \bf{0.265}  & \bf{0.031} & \bf{0.022}  & \bf{0.041}  & \bf{0.007}  & \bf{0.004}  & \bf{0.027}  & \bf{0.006}  & \bf{0.005}  \\
			     & (0.121)     & (0.096)     & (0.083)    & (0.241)     & (0.057)    & (0.043)     & (0.116)     & (0.046)     & (0.035)     & (0.100)     & (0.048)     & (0.037)     \\
			150  & \bf{0.475}  & \bf{0.483}  & \bf{0.489} & \bf{0.289}  & \bf{0.023} & \bf{0.014}  & \bf{0.027}  & \bf{0.002}  & \bf{0.001}  & \bf{0.013}  & \bf{0.002}  & \bf{0.001}  \\
			     & (0.118)     & (0.093)     & (0.080)    & (0.252)     & (0.042)    & (0.031)     & (0.095)     & (0.033)     & (0.028)     & (0.078)     & (0.036)     & (0.030)     \\
			300  & \bf{0.475}  & \bf{0.485}  & \bf{0.488} & \bf{0.121}  & \bf{0.027} & \bf{0.010}  & \bf{0.007}  & \bf{0.001}  & \bf{0.000}  & \bf{0.003}  & \bf{0.001}  & \bf{0.000}  \\ 
			     & (0.111)     & (0.090)     & (0.078)    & (0.161)     & (0.035)    & (0.022)     & (0.048)     & (0.024)     & (0.020)     & (0.037)     & (0.026)     & (0.021)     \\ \cline{1-13} 
		\end{tabular}
	\end{adjustbox}
\end{table}
\begin{table}[H]
	\addtocounter{table}{-1}
	\renewcommand{\thetable}{\arabic{table}c}
	\caption{Bias (SE) $\beta_2$}\label{T1c}
	\begin{adjustbox}{width=15cm,center} 
		\begin{tabular}{|c|rrr|rrr|rrr|rrr|}\cline{1-13} 
			& \multicolumn{3}{c|}{LS} & \multicolumn{3}{c|}{PC}   & \multicolumn{3}{c|}{BN} & \multicolumn{3}{c|}{QPC} \\ \cline{1-13} 
			\multicolumn{1}{|c|}{$n\ \backslash \ T$}   & \multicolumn{1}{c}{$6$} & \multicolumn{1}{c}{$9$} & \multicolumn{1}{c|}{$12$} &  \multicolumn{1}{c}{$6$} & \multicolumn{1}{c}{$9$} & \multicolumn{1}{c|}{$12$} &  \multicolumn{1}{c}{$6$} & \multicolumn{1}{c}{$9$} & \multicolumn{1}{c|}{$12$} &  \multicolumn{1}{c}{$6$} & \multicolumn{1}{c}{$9$} & \multicolumn{1}{c|}{$12$} \\ \cline{1-13} 
			30   & \bf{0.001}  & \bf{0.000}  & \bf{0.001} & \bf{-0.054} & \bf{-0.003}& \bf{-0.001} & \bf{-0.020} & \bf{-0.002} & \bf{-0.001} & \bf{-0.055} & \bf{-0.003} & \bf{-0.001} \\
			     & (0.142)     & (0.100)     & (0.088)    & (0.197)     & (0.095)    & (0.081)     & (0.199)     & (0.097)     & (0.082)     & (0.236)     & (0.104)     & (0.085)     \\
			60   & \bf{-0.001} & \bf{0.000}  & \bf{-0.001}& \bf{-0.047} & \bf{-0.004}& \bf{-0.001} & \bf{-0.009} & \bf{-0.002} & \bf{-0.001} & \bf{-0.016} & \bf{-0.004} & \bf{-0.001}  \\
		         & (0.096)     & (0.077)     & (0.065)    & (0.134)     & (0.075)    & (0.059)     & (0.118)     & (0.076)     & (0.059)     & (0.135)     & (0.081)     & (0.062)     \\
			90   & \bf{-0.001} & \bf{0.001}  & \bf{-0.001}& \bf{-0.097} & \bf{-0.002}& \bf{0.000}  & \bf{-0.006} & \bf{0.000}  & \bf{0.000}  & \bf{-0.008} & \bf{-0.001} & \bf{0.000}  \\
			     & (0.081)     & (0.062)     & (0.053)    & (0.141)     & (0.059)    & (0.048)     & (0.095)     & (0.060)     & (0.048)     & (0.104)     & (0.065)     & (0.051)     \\
			150  & \bf{0.000}  & \bf{0.000}  & \bf{0.000} & \bf{-0.012} & \bf{-0.002}& \bf{-0.001} & \bf{-0.003} & \bf{0.000}  & \bf{0.000}  & \bf{-0.003} & \bf{0.000}  & \bf{0.000}  \\
			     & (0.067)     & (0.048)     & (0.043)    & (0.137)     & (0.045)    & (0.039)     & (0.081)     & (0.045)     & (0.040)     & (0.089)     & (0.050)     & (0.042)     \\
			300  & \bf{0.000}  & \bf{0.000}  & \bf{0.000} & \bf{-0.046} & \bf{-0.002}& \bf{-0.001} & \bf{-0.002} & \bf{0.000}  & \bf{0.000}  & \bf{-0.002} & \bf{0.000}  & \bf{0.000}  \\ 
				 & (0.042)     & (0.035)     & (0.031)    & (0.087)     & (0.033)    & (0.028)     & (0.044)     & (0.033)     & (0.028)     & (0.049)     & (0.036)     & (0.031)     \\ \cline{1-13} 
		\end{tabular}
	\end{adjustbox}
\end{table}

Inspecting Table \ref{T1a}, the LS estimates of $\alpha$ appear to perform relatively well, which is expected since the model is not transformed in any way and the errors and factors are both drawn independently in each time period. The PC estimates of $\alpha$ on the other hand, suffer from a bias with fixed $T$ originating from the implicit transformation of the model to remove the factor term, which generates Nickell bias in the autoregressive coefficient. As expected, both the BN and the QPC estimates of $\alpha$ are unbiased as $n$ increases. For the coefficient $\beta_1$, the LS estimates are severely biased, with this bias being persistent irrespective of $n$ and $T$. For small $T$, the PC estimates are also biased, which stems from the heteroskedasticity of the errors in both the cross-section and across time. Only where both $n$ and $T$ are large does this bias diminish. Owing to the significant heteroskedasticity in the design, when both $n$ and $T$ are small, BN and QPC have sizeable biases - though smaller in magnitude than LS or PC. This bias diminishes rapidly as $n$ increases. Since $\bs{X}_2$ is neither dynamic, nor correlated with the factor term, estimates of $\beta_2$ generally perform well across all $n$ and $T$. Tables \ref{T2a} -- \ref{T2c} below present coverage probabilities of the estimators based on the asymptotic variance-covariance matrix, and with a nominal value of $95\%$.   

\begin{table}[H]
	\renewcommand{\thetable}{\arabic{table}a}
	\caption{Coverage $\alpha$ \%}\label{T2a}
	\begin{adjustbox}{width=15cm,center} 
		\begin{tabular}{|c|rrr|rrr|rrr|rrr|}\cline{1-13} 
			& \multicolumn{3}{c|}{LS} & \multicolumn{3}{c|}{PC}   & \multicolumn{3}{c|}{BN} & \multicolumn{3}{c|}{QPC} \\ \cline{1-13} 
			\multicolumn{1}{|c|}{$n\ \backslash \ T$}   & \multicolumn{1}{c}{$6$} & \multicolumn{1}{c}{$9$} & \multicolumn{1}{c|}{$12$} &  \multicolumn{1}{c}{$6$} & \multicolumn{1}{c}{$9$} & \multicolumn{1}{c|}{$12$} &  \multicolumn{1}{c}{$6$} & \multicolumn{1}{c}{$9$} & \multicolumn{1}{c|}{$12$} &  \multicolumn{1}{c}{$6$} & \multicolumn{1}{c}{$9$} & \multicolumn{1}{c|}{$12$} \\ \cline{1-13} 
			30   & 85.35 & 85.30 & 86.65 & 60.32 & 83.73 & 88.69 & 77.53 & 88.66 & 89.84 & 45.21 & 72.39 & 70.86 \\
			60   & 85.47 & 86.74 & 87.31 & 53.57 & 79.59 & 86.39 & 85.32 & 91.16 & 92.68 & 67.83 & 74.33 & 78.90 \\
			90   & 86.47 & 87.14 & 87.61 & 30.72 & 76.51 & 83.49 & 87.40 & 91.98 & 93.08 & 72.46 & 80.35 & 81.91 \\
			150  & 86.99 & 86.26 & 88.01 & 22.63 & 71.12 & 79.61 & 88.83 & 93.29 & 93.34 & 71.22 & 86.40 & 85.66 \\
			300  & 84.46 & 87.16 & 87.70 & 27.85 & 62.05 & 72.27 & 91.22 & 93.63 & 93.91 & 82.59 & 88.49 & 90.03 \\  \cline{1-13} 
		\end{tabular}
	\end{adjustbox}
\end{table}

\begin{table}[H]
	\addtocounter{table}{-1}
	\renewcommand{\thetable}{\arabic{table}b}
	\caption{Coverage $\beta_1$ \%}\label{T2b}
	\begin{adjustbox}{width=15cm,center} 
		\begin{tabular}{|c|rrr|rrr|rrr|rrr|}\cline{1-13} 
			& \multicolumn{3}{c|}{LS} & \multicolumn{3}{c|}{PC}   & \multicolumn{3}{c|}{BN} & \multicolumn{3}{c|}{QPC} \\ \cline{1-13} 
			\multicolumn{1}{|c|}{$n\ \backslash \ T$}   & \multicolumn{1}{c}{$6$} & \multicolumn{1}{c}{$9$} & \multicolumn{1}{c|}{$12$} &  \multicolumn{1}{c}{$6$} & \multicolumn{1}{c}{$9$} & \multicolumn{1}{c|}{$12$} &  \multicolumn{1}{c}{$6$} & \multicolumn{1}{c}{$9$} & \multicolumn{1}{c|}{$12$} &  \multicolumn{1}{c}{$6$} & \multicolumn{1}{c}{$9$} & \multicolumn{1}{c|}{$12$} \\ \cline{1-13} 
			30   & 00.84 & 00.02 & 00.00 & 53.68 & 74.19 & 80.50 & 64.39 & 81.51 & 84.43 & 61.75 & 82.86 & 85.62 \\
			60   & 00.13 & 00.00 & 00.00 & 62.17 & 80.23 & 82.98 & 77.18 & 88.67 & 91.07 & 74.38 & 86.42 & 90.40 \\
			90   & 00.07 & 00.00 & 00.00 & 35.07 & 81.29 & 84.92 & 80.65 & 90.98 & 93.16 & 80.15 & 87.51 & 92.01 \\
			150  & 00.03 & 00.00 & 00.00 & 33.31 & 81.52 & 87.90 & 82.75 & 92.56 & 93.79 & 79.76 & 89.44 & 92.03 \\
			300  & 00.00 & 00.00 & 00.00 & 44.95 & 73.44 & 88.59 & 89.63 & 93.00 & 94.01 & 82.85 & 90.88 & 92.79 \\  \cline{1-13} 
		\end{tabular}
	\end{adjustbox}
\end{table}

\begin{table}[H]
	\addtocounter{table}{-1}
	\renewcommand{\thetable}{\arabic{table}b}
	\caption{Coverage $\beta_2$ \%}\label{T2c}
	\begin{adjustbox}{width=15cm,center} 
		\begin{tabular}{|c|rrr|rrr|rrr|rrr|}\cline{1-13} 
			& \multicolumn{3}{c|}{LS} & \multicolumn{3}{c|}{PC}   & \multicolumn{3}{c|}{BN} & \multicolumn{3}{c|}{QPC} \\ \cline{1-13} 
			\multicolumn{1}{|c|}{$n\ \backslash \ T$}   & \multicolumn{1}{c}{$6$} & \multicolumn{1}{c}{$9$} & \multicolumn{1}{c|}{$12$} &  \multicolumn{1}{c}{$6$} & \multicolumn{1}{c}{$9$} & \multicolumn{1}{c|}{$12$} &  \multicolumn{1}{c}{$6$} & \multicolumn{1}{c}{$9$} & \multicolumn{1}{c|}{$12$} &  \multicolumn{1}{c}{$6$} & \multicolumn{1}{c}{$9$} & \multicolumn{1}{c|}{$12$} \\ \cline{1-13} 
			30   & 86.54 & 84.87 & 85.91 & 84.78 & 93.09 & 93.49 & 86.74 & 92.95 & 93.38 & 75.79 & 90.91 & 92,29 \\
			60   & 84.79 & 86.30 & 86.42 & 81.87 & 92.86 & 93.62 & 88.34 & 92.94 & 93.48 & 80.25 & 89.33 & 92.23 \\
			90   & 85.72 & 86.17 & 87.23 & 66.25 & 93.29 & 94.16 & 90.55 & 93.22 & 94.16 & 85.79 & 88.76 & 92.70 \\
			150  & 88.13 & 86.56 & 87.71 & 53.81 & 93.17 & 93.92 & 91.03 & 93.17 & 93.86 & 83.23 & 90.15 & 92.98 \\
			300  & 83.71 & 87.26 & 87.41 & 74.34 & 93.56 & 93.55 & 91.89 & 93.50 & 93.92 & 84.16 & 91.66 & 93.07 \\  \cline{1-13} 
		\end{tabular}
	\end{adjustbox}
\end{table}
For $\alpha$ the coverage of LS remains consistently below its nominal value, while for PC it decreases with fixed $T$. In the case of the latter, this decrease in coverage is expected due to the fixed $T$ bias, with coverage only improving when both $n$ and $T$ increase. In contrast, the coverage of BN and QPC readily improve as $n$ increases, with $T$ fixed or $T$ increasing slowly. The story is similar for $\beta_1$ in Table \ref{T2b}. The coverage of LS is incredibly poor, presenting near $0$ across all $n,T$ values. The coverage of PC is also poor with either $n$ or $T$ small, and improves only as both of these increase. BN and QPC present poor coverage with both $n$ and $T$ small, yet these rapidly improve as $n$ increases. When comparing the performance of BN and QPC, it is, in general, the case that BN outperforms QPC. This is a consequence of the fact that, while omitting a time period, BN uses only $2$ factors in estimation, whereas QPC uses $3$, with the extra factor being present to control for a possibly endogenous initial condition. Clearly, including an additional factor in estimation has a noticeable impact on the efficiency of the estimator in finite samples, therefore it is useful to apply the eigenvalue ratio test described in Section \ref{estfac} to uncover the appropriate number of factors to use in estimation.

\begin{table}[H]
	\renewcommand{\thetable}{3}
	\caption{Number of Factors Chosen QPC\ \%}	
	\begin{adjustbox}{width=10cm,center} \label{T23g}
		\begin{tabular}{|c|rrr|rrr|rrr|}\cline{1-7} 
			& \multicolumn{3}{c|}{EigR = 2} & \multicolumn{3}{c|}{EigR = 3} \\ \cline{1-7} 
			\multicolumn{1}{|c|}{$n\ \backslash \ T$}   & \multicolumn{1}{c}{$6$} & \multicolumn{1}{c}{$9$} & \multicolumn{1}{c|}{$12$} & \multicolumn{1}{c}{$6$} & \multicolumn{1}{c}{$9$} & \multicolumn{1}{c|}{$12$} \\ \cline{1-7} 
			30   & 27.26 & 41.36 & 47.15 & 30.58 & 08.66 & 05.28 \\
			60   & 40.74 & 60.39 & 70.11 & 11.86 & 01.01 & 00.35 \\
			90   & 45.00 & 70.14 & 77.21 & 07.67 & 00.29 & 00.11 \\
			150  & 55.09 & 79.98 & 89.01 & 03.52 & 00.06 & 00.00 \\
			300  & 73.36 & 85.15 & 93.62 & 00.02 & 00.00 & 00.00 \\ \cline{1-7} 
		\end{tabular}
	\end{adjustbox}
\end{table}

Table \ref{T23g} presents the percentage of times that the number of factors is chosen to be either $2$ or $3$ when applying the modified eigenvalue ratio test (EigR) described in Section \ref{estfac} to the QPC residuals. Only in the smallest sample size, $n = 30$, $T=6$, is the number of factors chosen to be $3$. In all other cases the percentage of times $3$ factors is chosen is very small. Indeed, this number declines rapidly as either $n$ or $T$ grows, while the number of times $2$ is chosen increases. This suggests that the impact of the initial condition becomes negligible as either dimension of the panel grows. In light of this, Tables \ref{T23a} and \ref{T23b} below present bias and coverages for QPC with $R = 2$. Comparing these results to those presented previously, these estimates are generally better than both BN and QPC with an additional factor. This is unsurprising since it uses both the lower number of factors, and retains an additional time period when compared to BN. However, BN still outperforms QPC when it comes to the autoregressive parameter $\alpha$.
\begin{table}[H]
	\renewcommand{\thetable}{\arabic{table}a}
	\caption{Bias (SE), QPC with $R = 2$}
	\begin{adjustbox}{width=13cm,center} \label{T23a}
		\begin{tabular}{|c|rrr|rrr|rrr|rrr|}\cline{1-10} 
			& \multicolumn{3}{c|}{$\alpha$} & \multicolumn{3}{c|}{$\beta_1$} & \multicolumn{3}{c|}{$\beta_2$} \\ \cline{1-10} 
			\multicolumn{1}{|c|}{$n\ \backslash \ T$}   & \multicolumn{1}{c}{$6$} & \multicolumn{1}{c}{$9$} & \multicolumn{1}{c|}{$12$} & \multicolumn{1}{c}{$6$} & \multicolumn{1}{c}{$9$} & \multicolumn{1}{c|}{$12$} & \multicolumn{1}{c}{$6$} & \multicolumn{1}{c}{$9$} & \multicolumn{1}{c|}{$12$}  \\ \cline{1-10} 
            30   & \bf{-0.034}  & \bf{-0.004} & \bf{-0.002} & \bf{0.126}  & \bf{0.057} & \bf{0.047}  & \bf{-0.013} & \bf{-0.002} & \bf{-0.001} \\
                 & (0.150)      & (0.054)     & (0.041)     & (0.211)     & (0.108)    & (0.088)     & (0.171)     & (0.097)     & (0.083)     \\
			60   & \bf{-0.008}  & \bf{-0.003} & \bf{-0.001} & \bf{0.040}  & \bf{0.012} & \bf{0.009}  & \bf{-0.006} & \bf{-0.001} & \bf{-0.001} \\
			     & (0.072)      & (0.039)     & (0.029)     & (0.111)     & (0.058)    & (0.045)     & (0.103)     & (0.071)     & (0.058)     \\
			90   & \bf{-0.006}  & \bf{-0.001} & \bf{-0.001} & \bf{0.027}  & \bf{0.006} & \bf{0.005}  & \bf{-0.004} & \bf{0.000}  & \bf{0.000}  \\
			     & (0.058)      & (0.032)     & (0.025)     & (0.087)     & (0.043)    & (0.036)     & (0.084)     & (0.057)     & (0.048)     \\
			150  & \bf{-0.002}  & \bf{-0.001} & \bf{0.000}  & \bf{0.013}  & \bf{0.002} & \bf{0.001}  & \bf{-0.001} & \bf{0.000}  & \bf{0.000}  \\
			     & (0.048)      & (0.024)     & (0.020)     & (0.064)     & (0.032)    & (0.027)     & (0.069)     & (0.044)     & (0.038)     \\ 
			300  & \bf{-0.001}  & \bf{-0.001} & \bf{0.000}  & \bf{0.003}  & \bf{0.001} & \bf{0.000 } & \bf{-0.001} & \bf{0.000}  & \bf{0.000}  \\
			     & (0.027)      & (0.018)     & (0.015)     & (0.034)     & (0.024)    & (0.020)     & (0.039)     & (0.034)     & (0.028)     \\ \cline{1-10}
		\end{tabular}
	\end{adjustbox}
\end{table}                                                                                                                                                                   

\begin{table}[H]
	\addtocounter{table}{-1}
	\renewcommand{\thetable}{\arabic{table}b}
	\caption{Bias (SE), QPC with $R = 2$}
	\begin{adjustbox}{width=13cm,center} \label{T23b}
		\begin{tabular}{|c|rrr|rrr|rrr|rrr|}\cline{1-10} 
			& \multicolumn{3}{c|}{$\alpha$} & \multicolumn{3}{c|}{$\beta_1$} & \multicolumn{3}{c|}{$\beta_2$} \\ \cline{1-10} 
			\multicolumn{1}{|c|}{$n\ \backslash \ T$}   & \multicolumn{1}{c}{$6$} & \multicolumn{1}{c}{$9$} & \multicolumn{1}{c|}{$12$} & \multicolumn{1}{c}{$6$} & \multicolumn{1}{c}{$9$} & \multicolumn{1}{c|}{$12$}  & \multicolumn{1}{c}{$6$} & \multicolumn{1}{c}{$9$} & \multicolumn{1}{c|}{$12$} \\ \cline{1-10} 
			30   & 59.75 & 78.99 & 76.18 & 67.00 & 76.62 & 78.19 & 88.19 & 92.71 & 93,31 \\
			60   & 82.18 & 84.03 & 86.08 & 81.02 & 89.55 & 91.50 & 90.95 & 93.61 & 94.16 \\
			90   & 83.72 & 87.77 & 86.74 & 84.62 & 91.37 & 93.03 & 92.85 & 93.25 & 94.01 \\
			150  & 84.75 & 91.32 & 89.04 & 87.86 & 93.05 & 93.80 & 92.26 & 93.36 & 93.89 \\
			300  & 92.34 & 92.39 & 92.11 & 90.95 & 93.42 & 94.06 & 92.39 & 93.47 & 94.02 \\ \cline{1-10}
		\end{tabular}
	\end{adjustbox}
\end{table}

\section{Conclusion}\label{concl}
In conclusion, this paper introduces a new method to estimate linear panel data models with interactive fixed effects designed for situations where $T$ is small relative to $n$, or indeed may be fixed. By transforming the model and then applying the PC estimator of \cite{bai_panel_2009}, the approach proposed in this paper is shown to deliver $\sqrt{n}$ consistent estimates of regression slope coefficients with $T$ fixed which are, moreover,  asymptotically unbiased in the presence of cross-sectional dependence, serial dependence, and with the inclusion of dynamic regressors. This contrasts starkly with the usual case where the PC estimator generally delivers biased and inconsistent estimates with $T$ fixed. Several other consequences of this approach are also discussed, particularly the ability to apply other inferential procedures designed for the large $n$, large $T$ setting to the transformed model. This is illustrated by modifying to the eigenvalue ratio test of \cite{horenstein} to render it applicable in the present setting. 

There are two natural extensions to this paper, both of which are currently in progress. The first is to notice that the estimator proposed in this paper can be obtained as a marginal likelihood associated with a maximal invariant statistic under the group of transformations \eqref{group}. Using the full likelihood of the maximal invariant may potentially lead to improved estimation of the autoregressive parameter, as has been shown in a similar context by \cite{moriera_dynamic}. The second extension is to incorporate more general predetermined regressors which are intuitively difficult to handle in this framework. 

The results and perspective introduced in this paper also suggest several other avenues for future research. For example, the transformation introduced in this paper, or similar, might be applied to other inferential procedures designed for a large panels, potentially rendering them amenable to the fixed $T$ setting. Finally, it is worth stressing that arguably the most powerful concept developed in this research is the idea that multi-dimensional nuisance parameters may be removed from one (or possibly several) dimensions, by reducing the model into a lower dimensional subspace. This, really, is what lies at the heart of this paper and may well prove to be fruitful in other applications.